\documentclass[english,aps,prc,nofootinbib,superscriptaddress,twocolumn,letter]{revtex4}
\usepackage[latin9]{inputenc}
\usepackage{hyperref}
\hypersetup{
  colorlinks=true,        
  linkcolor=blue,         
  citecolor=cyan,         
}
\usepackage{breakurl}
\usepackage{graphicx}
\usepackage{epsf}
\usepackage{epsfig}
\usepackage{amssymb,amsmath}
\usepackage[usenames]{color}
\usepackage{amssymb}
\usepackage{times}
\usepackage{comment}

\newcommand\snn{\sqrt{s_\text{NN}}}

\newcommand\ycm{y_{\textrm{CM}}}



\begin{document}

\preprint{This line only printed with preprint option}

\title{{\color{black} Longitudinal distribution of initial energy density and directed flow of charged particles in relativistic heavy-ion collisions}}

\author{Ze-Fang Jiang}
\email{jiangzf@mails.ccnu.edu.cn}
\affiliation{Institute of Particle Physics and Key Laboratory of Quark and Lepton Physics (MOE), Central China Normal University, Wuhan, Hubei, 430079, China}
\affiliation{Department of Physics and Electronic-Information Engineering, Hubei Engineering University, Xiaogan, Hubei, 432000, China}

\author{Shanshan Cao}
\email{shanshan.cao@sdu.edu.cn}
\affiliation{Institute of Frontier and Interdisciplinary Science, Shandong University, Qingdao, Shandong 266237, China}

\author{Xiang-Yu Wu}
\affiliation{Institute of Particle Physics and Key Laboratory of Quark and Lepton Physics (MOE), Central China Normal University, Wuhan, Hubei, 430079, China}

\author{C. B. Yang}
\affiliation{Institute of Particle Physics and Key Laboratory of Quark and Lepton Physics (MOE), Central China Normal University, Wuhan, Hubei, 430079, China}

\author{Ben-Wei Zhang}
\affiliation{Institute of Particle Physics and Key Laboratory of Quark and Lepton Physics (MOE), Central China Normal University, Wuhan, Hubei, 430079, China}

\begin{abstract}
We study the origin of the directed flow of charged particles produced in relativistic heavy-ion collisions. Three different initial conditions, Boz$\dot{\textrm{e}}$k-Wyskiel, CCNU and Shen-Alzhrani, of energy density distributions are coupled to the (3+1)-dimensional viscous hydrodynamic model CLVisc, and their effects on the development of the anisotropic medium geometry, pressure gradient and radial flow are systematically compared. By comparing to experimental data at both RHIC and LHC, we find that the directed flow provides a unique constraint on the tilt of the initial medium profile in the plane spanned by the impact parameter and space-time rapidity. Within mid-rapidity, the counter-clockwise tilt is shown to be a crucial source of the positive/negative force by the pressure gradient along the impact parameter ($x$) direction at backward/forward rapidity, which drives a negative slope of the $x$ component of the medium flow velocity with respect to rapidity, and in the end the same feature of the charged particle directed flow.
\end{abstract}
\maketitle
\date{\today}

\section{Introduction}
\label{emsection1}

Heavy-ion collisions at the BNL Relativistic Heavy-Ion Collider (RHIC) and the CERN Large Hadron Collider (LHC) suggest that a hot and dense nuclear matter, known as quark-gluon plasma (QGP), is formed in the reaction region.
The strong collective flow, such as the elliptic flow ($v_2$), of the observed hadrons in various collision systems~\cite{PHENIX:2003qra,ALICE:2010suc,CMS:2012zex} is one of the most important signatures of the strongly interacting nature of the QGP. It has been successfully described by relativistic hydrodynamic models~\cite{Ollitrault:1992bk,Rischke:1995ir,Sorge:1996pc,Bass:1998vz,Aguiar:2001ac,Shuryak:2003xe,Gyulassy:2004zy,Broniowski:2007ft,Andrade:2008xh,Hirano:2009ah,Schenke:2010rr,Qiu:2011iv,Heinz:2013th,Huovinen:2013wma,Gale:2013da,Bozek:2013uha,Qin:2013bha,Dusling:2015gta,Romatschke:2017ejr,Weller:2017tsr,Zhao:2020wcd}, and the specific shear viscosity extracted from the model-to-data comparison is shown small~\cite{Song:2010mg,Bernhard:2019bmu}.

The first-order Fourier coefficient of the azimuthal distribution of hadrons, also known as the rapidity-odd directed flow ($v_{1}$)~\cite{Voloshin:1994mz,Bilandzic:2010jr}, is among the earliest observables for studying the collectivity in nuclear collisions at different energies~\cite{Gyulassy:1981nq,Gustafsson:1984ka,Lisa:2000ip}.
Since the commencement of relativistic heavy-ion collisions, it has been widely studied in RHIC and LHC experiments as well~\cite{STAR:2004jwm,STAR:2014clz,STAR:2017okv,STAR:2019clv,ALICE:2019sgg,STAR:2019vcp}. Model calculations suggest that directed flow is initiated during the passage time of the two colliding nuclei, whose typical time scale is $2R/\gamma$ with $R$ and $\gamma$ being the nuclear radius and Lorentz contraction factor respectively~\cite{Gyulassy:1981nq,Sorge:1996pc,Singha:2016mna}. This could be earlier than the development of elliptic flow. Therefore, $v_1$ has been considered a sensitive probe of the fireball size and nucleon flow at the initial stage~\cite{Ollitrault:1992bk,Voloshin:1994mz,Nara:2016phs,Chatterjee:2017ahy,Singha:2016mna,Zhang:2018wlk,Guo:2017mkf}. There are various sources contributing to the directed flow. It has been proposed that the size and sign of $v_1$ could depend on the deformation of the initial medium geometry, the baryon current, the equation of state of nuclear matter and also hadronic rescatterings~\cite{Adil:2005qn,Bozek:2010bi,Chen:2019qzx,Shen:2020jwv,Ryu:2021lnx,Chatterjee:2017ahy,Chatterjee:2018lsx,Beraudo:2021ont}, although their quantitative relative contributions are still open questions.

To investigate how the initial geometric asymmetry is transformed to the final hadron $v_1$, various initialization methods have been developed and coupled to hydrodynamic model calculations. Among different parametrizations of the longitudinal structure of the initial fireball, one of the most frequently applied approaches is the Boz$\dot{\textrm{e}}$k-Wyskiel parametrization proposed in Ref.~\cite{Bozek:2010bi}. It takes into account the asymmetry along the impact parameter direction ($x$) at different space-time rapidity ($\eta_s$) which generates a counter-clockwise tilt of the initial fireball in the $x$-$\eta_s$ plane. This initialization method has been found successful in understanding the directed flow of not only the soft hadron, but also the heavy flavor mesons ~\cite{Chatterjee:2018lsx,Beraudo:2021ont,Chatterjee:2017ahy,Oliva:2020doe} after coupling the heavy quark evolution model to the titled QGP medium. Following the idea of Boz$\dot{\textrm{e}}$k-Wyskiel, we developed an alternative initialization ansatz (CCNU parametrization) of the longitudinal distribution of the nuclear matter in an earlier study~\cite{Jiang:2021foj}. It grasps the key feature of the tilted medium geometry and is able to describe the charged particle $v_1$ at RHIC and LHC.
Recently, an additional collision geometry-based 3-dimensional (3D) initial condition (Shen-Alzhrani parametrization) is developed in Refs.~\cite{Shen:2020jwv,Ryu:2021lnx}, which incorporates the Bjorken flow in the longitudinal direction (same as the Boz$\dot{\textrm{e}}$k-Wyskiel and CCNU parametrizations) and ensures the conservation of local energy and momentum, and provides a satisfactory description of the $v_1$ of $\pi^{+}$ at RHIC. Therefore, it is of great interest to conduct a detailed comparison between these different initialization approaches within a uniform QGP evolution framework, and identify the main features of the initial geometry that lead to the final state hadron $v_1$ we observe.

In this work, the abovementioned systematical comparison between the three initial conditions -- Boz$\dot{\textrm{e}}$k-Wyskiel, CCNU and Shen-Alzhrani -- is performed using the (3+1)-D viscous hydrodynamic model CLVisc~\cite{Pang:2016igs,Pang:2018zzo,Wu:2018cpc}. We investigate the correlation between the longitudinal structure of the initial fireball and the directed flow of the final state charged particles. Our calculation indicates the counter-clockwise tilt of the initial energy density profile yields an increasing/decreasing average pressure gradient $-\langle \partial_x P\rangle$ from zero with respect to time at backward/forward space-time rapidity within $|\eta_s|<2$. This further leads to a negative slope of the average QGP flow velocity $\langle v_x \rangle$ with respect to $\eta_s$, and in the end the same behavior of $v_1$ {\it vs.} $\eta$ for the final-state charged particles, which is consistent with the experimental observations in $\snn=200$~GeV Au+Au collisions at RHIC~\cite{Abelev:2008jga} and $\snn=$ 2.76~TeV Pb+Pb collisions at LHC~\cite{Abelev:2013cva}.

This article is organized as follows. In Sec.~\ref{v1section2}, we will discuss three different initialization methods of the rapidity-dependent energy density distribution in heavy-ion collisions and their impacts on the pressure gradient and flow velocity during the hydrodynamic expansion of the QGP.  
In Sec.~\ref{v1section3}, we will present the charged particle directed flow from our hydrodynamic calculation and investigate its dependence on the initial condition. In the end, we summarize and discuss future improvements in Sec.~\ref{v1section4}.

\section{The model framework}
\label{v1section2}
\subsection{Parametrizations of the initial energy density profile}
\label{v1subsect2}
For the purpose of investigating the dependence of the final-state directed flow on the initial-state geometry of nuclear matter, in this section, we construct three different initial energy density distributions based on pioneer studies~\cite{Bozek:2010bi,Bozek:2011ua,Shen:2020jwv,Ryu:2021lnx,Jiang:2021foj}.
Their impacts on the time evolution of the pressure gradient and flow velocity of the QGP will be then explored using a (3+1)-D hydrodynamic model.

The nucleus thickness function $T(x,y)$ from the Woods-Saxon distribution is
\begin{equation}
\begin{aligned}
T(x,y)=\int_{-\infty}^{\infty}dz\frac{n_{0}}{1+e^{(\sqrt{x^{2}+y^{2}+z^{2}}-R)/d}},
\label{eq:thicknessf}
\end{aligned}
\end{equation}
where $n_{0}$ is the average nuclear density, $d$ is the diffusiveness parameter, $x,~y,~z$ are the space coordinates and $R$ is the radius of the nuclear Fermi distribution, which depends on the nucleus species. The parameters used for Au and Pb in the present study are listed in Tab.~\ref{t:parameters}.
\begin{table}[!h]
\begin{center}
\begin{tabular}{l c c c c}
\hline\hline
Nucleus     & $A$        & $n_{0}$ [1/fm$^{3}$]      & $R$~[fm]    & $d$ [fm]    \\ \hline
Au          & 197        & 0.17                      & 6.38      & 0.546       \\
Pb          & 208        & 0.17                      & 6.62      & 0.535       \\
\hline\hline
\end{tabular}
\caption{\label{t:parameters} Parameters used in the Woods-Saxon distribution for Au and Pb nuclei~\cite{Loizides:2017ack}.}
\end{center}
\end{table}

Consider two nuclei propagate along $\pm \hat{z}$ and collide with the impact parameter $\mathbf{b}$. Their thickness function may be written as
\begin{equation}
\begin{aligned}
T_{+}(\mathbf{x}_\text{T})=T(\mathbf{x}_\text{T}-\mathbf{b}/2),~~~~T_{-}(\mathbf{x}_\text{T})=T(\mathbf{x}_\text{T}+\mathbf{b}/2)
\label{eq:t+}
\end{aligned}
\end{equation}
respectively, where $\mathbf{x}_\text{T}=(x,y)$ is the transverse plane coordinate. The density distributions of participant nucleons from the two nuclei are then
\begin{equation}
\begin{aligned}
T_{1}(\mathbf{x}_\text{T})=T_{+}(\mathbf{x}_\text{T})\left\{1-\left[1-\frac{\sigma_\text{NN} T_{-}(\mathbf{x}_\text{T})}{A}\right]^{A}\right\},
\label{eq:t1}
\end{aligned}
\end{equation}
\begin{equation}
\begin{aligned}
T_{2}(\mathbf{x}_\text{T})=T_{-}(\mathbf{x}_\text{T})\left\{1-\left[1-\frac{\sigma_\text{NN} T_{+}(\mathbf{x}_\text{T})}{A}\right]^{A}\right\},
\label{eq:t2}
\end{aligned}
\end{equation}
where $A$ is the mass number of the colliding nuclei, $\sigma_\text{NN}$ is the inelastic nucleon-nucleon scattering cross section.
The collision centrality classes are determined by the impact parameter $\mathbf{b}$~\cite{Loizides:2017ack}.

The right/left-moving wounded nucleons are expected to produce more particles at forward/backward rapidity, respectively.
This effect can be obtained by introducing the rapidity-dependent deformation into the weight function $W_\text{N}$ of wounded nucleons.

\underline{\textbf{\emph{Case (A)}}} Boz$\dot{\textrm{e}}$k-Wyskiel parametrization.

Following the Boz$\dot{\textrm{e}}$k-Wyskiel parametrization of the tilted initial condition~\cite{Bozek:2010bi,Bozek:2011ua}, two piecewise functions are used to construct the wounded nucleon weight function $W_\text{N}$ as:
\begin{equation}
\begin{aligned}
W_\text{N}(x,y,\eta_{s})=2\left[T_{1}(x,y)f_{+}(\eta_{s}) + T_{2}(x,y)f_{-}(\eta_{s})\right],
\label{eq:wnpl}
\end{aligned}
\end{equation}
with
$$ f_{+}(\eta_{s})=\left\{
\begin{aligned}
&0,                                 && \eta_{s} < -\eta_{m}, \\
&\frac{\eta_{s}+\eta_{m}}{2\eta_{m}},  && -\eta_{m} \leq \eta_{s} \leq\eta_{m},\\
&1,                                 & &\eta_{s} > \eta_{m},
\end{aligned}
\right.
$$
and
$$ f_{-}(\eta_{s})=\left\{
\begin{aligned}
&1,                                 && \eta_{s} < -\eta_{m}, \\
&\frac{-\eta_{s}+\eta_{m}}{2\eta_{m}},  && -\eta_{m} \leq \eta_{s} \leq\eta_{m},\\
&0,                                 & &\eta_{s} > \eta_{m},
\end{aligned}
\right.
$$
in which $\eta_{s}$ is the longitudinal space-time rapidity,
$\eta_{m}$ defines the range of rapidity correlations and affects the relative contribution from forward and backward participating nucleons.

\underline{\textbf{\emph{Case (B)}}} CCNU parametrization.

In our earlier study~\cite{Jiang:2021foj}, a monotonic function was introduced to describe the imbalance between the forward and backward regions, resulting in a tilted initial condition. The weight function of wounded nucleons $W_\text{N}$ is defined as,
\begin{equation}
\begin{aligned}
W_\text{N}(x,y,\eta_{s})=&[T_{1}(x,y)+T_{2}(x,y)]\\
+&H_{t}[T_{1}(x,y)-T_{2}(x,y)]\tan\left(\frac{\eta_{s}}{\eta_{t}}\right),
\label{eq:mnccnu}
\end{aligned}
\end{equation}
where the parameter $H_{t}$ reflects the overall strength of imbalance between particle emission in the forward and backward rapidities along the direction of the impact parameter, while $\tan (\eta_{s}/\eta_{t})$ generates the deformation of the initial energy density distribution along the rapidity direction. A fixed parameter of $\eta_{t}=8.0$ is used for all the collision systems investigated in the present work.

The initial energy density $\varepsilon(x,y,\eta_{s})$ for both \emph{\textbf{Case (A)}} and \emph{\textbf{Case (B)}} is then given by~\cite{Pang:2018zzo}
\begin{equation}
\begin{aligned}
\varepsilon(x,y,\eta_{s})=K \cdot W(x,y,\eta_{s}) \cdot H(\eta_{s}),
\label{eq:ekw}
\end{aligned}
\end{equation}
where $K$ is an overall normalization factor that will be determined by the particle yield in different collision systems, and $W(x,y,\eta_{s})$ is the total weight function defined as
\begin{equation}
\begin{aligned}
W(x,y,\eta_{s})=\frac{(1-\alpha)W_\text{N}(x,y,\eta_{s})+\alpha n_\text{BC}(x,y)}{\left[(1-\alpha)W_\text{N}(0,0,0)+\alpha n_\text{BC}(0,0)\right]|_{\mathbf{b}=0}}.
\label{eq:wneta}
\end{aligned}
\end{equation}
Here, $\alpha$ is known as the collision hardness parameter and $n_\text{BC}(x,y)$ is the number of binary (hard) collisions given by
\begin{equation}
\begin{aligned}
n_\text{BC}(x,y)=\sigma_\text{NN}T_{+}(x,y)T_{-}(x,y).
\label{eq:nbc}
\end{aligned}
\end{equation}
In Eq.~(\ref{eq:ekw}), a function
\begin{equation}
\begin{aligned}
H(\eta_{s})=\exp\left[-\frac{(|\eta_{s}|-\eta_{w})^{2}}{2\sigma^{2}_{\eta}}\theta(|\eta_{s}|-\eta_{w}) \right]
\label{eq:heta}
\end{aligned}
\end{equation}
is introduced in order to describe the plateau structure of the rapidity distribution of emitted hadrons, in which $\eta_{w}$ determines the width of the central rapidity plateau while $\sigma_{\eta}$ determines the width (speed) of the Gaussian decay outside the plateau region~\cite{Pang:2018zzo}. Model parameters -- $K$, $\eta_w$ and $\sigma_\eta$ -- will be summarized in Tab.~\ref{t:modelparameters} soon.

\underline{\textbf{\emph{Case (C)}}} Shen-Alzhrani parametrization.

A third parametrization of the $\eta_s$-dependent initial condition was proposed in Refs.~\cite{Shen:2020jwv,Ryu:2021lnx}, which ensures the local energy-momentum conservation during converting the two colliding nuclei into the energy density profile of the produced nuclear medium. The local invariant mass $M(x,y)$ and the center-of-mass rapidity $y_{\text{CM}}$ are respectively defined as,
\begin{equation}
\begin{aligned}
M(x,y) = m_\text{N}\sqrt{T_{1}^{2}+T_{2}^{2}+2T_{1}T_{2}\textrm{cosh}(2y_{\textrm{beam}})},
\label{eq:mxy}
\end{aligned}
\end{equation}
\begin{equation}
\begin{aligned}
y_{\textrm{CM}}(x,y) = \textrm{arctanh}\left[\frac{T_{1}-T_{2}}{T_{1}+T_{2}}\textrm{tanh}(y_{\textrm{beam}})\right],
\label{eq:ycm}
\end{aligned}
\end{equation}
where $y_{\textrm{beam}}=\textrm{arccosh}(\sqrt{s_\text{NN}}/2m_\text{N})$ is the rapidity of each nucleon inside the colliding nuclei, and $m_\text{N}$ is the nucleon mass.

The local energy density profile is then modeled as~\cite{Shen:2020jwv},
\begin{equation}
\begin{aligned}
\varepsilon(x,y,\eta_{s};&\ycm)=K \cdot\mathcal{N}_{e}(x,y)\\
&\times \exp{\Big [}-\frac{(|\eta_{s}-(\ycm-y_{L})|-\eta_{w})^{2}}{2\sigma^{2}_{\eta}} \\
&\times \theta(|\eta_{s}-(\ycm-y_{L})|-\eta_{w}){\Big]},
\label{eq:eqosu1}
\end{aligned}
\end{equation}
in which $K$ is the overall normalization factor and $y_{L}=f\ycm$ with $f \in [0,1]$. Here, the transverse density distribution $\mathcal{N}_{e}$ is determined by the local invariant mass $M(x,y)$ as
\begin{equation}
\begin{aligned}
\mathcal{N}_{e}(x,y) =  \frac{M(x,y)}{{\color{black}M(0,0)}\left[2\sinh(\eta_{w})+\sqrt{\frac{\pi}{2}}\sigma_{\eta}e^{\sigma^{2}_{\eta}/2}C_{\eta}\right]},
\label{eq:Nxy}
\end{aligned}
\end{equation}
\begin{equation}
\begin{aligned}
C_{\eta} = e^{\eta_{w}}\textrm{erfc}\left(-\sqrt{\frac{1}{2}}\sigma_{\eta}\right)+e^{-\eta_{w}}\textrm{erfc}\left(\sqrt{\frac{1}{2}}\sigma_{\eta}\right),
\end{aligned}
\end{equation}
where erfc($x$) is the complementary error function.

In Tab.~\ref{t:modelparameters}, we summarize the common parameters that are shared between our three model setups, Case(A), (B) and (C), including the initial time of the hydrodynamic evolution ($\tau_0$), overall normalization factor ($K$), collision hardness parameter ($\alpha$) [used in Case (A) and (B)], nucleon-nucleon inelastic cross section ($\sigma_\text{NN}$), rapidity plateau width ($\eta_w$) and the width of the Gaussian decay ($\sigma_\eta$). They are tuned to provide a reasonable description of the charged hadron pseudo-rapidity distributions in the most central collisions~\cite{Pang:2018zzo}, as will be shown in Figs.~\ref{f:auau200dndeta} and~\ref{f:pbpb2760dedeta}. Note that in order to compare the three parametrizations above in the same hydrodynamic evolution framework (CLVisc in this work), the parameters tuned here could be different from the original Boz$\dot{\textrm{e}}$k-Wyskiel work for Case (A) and Shen-Alzhrani work for Case (C). The parameter designed for each specific model -- $\eta_m$ for Case (A), $H_t$ for Case (B) and $f$ for Case (C) -- will be discussed later when we compare its corresponding model to the charged particle $v_1$ data.

\begin{table}[!h]
\begin{center}
\begin{tabular}{ |c| c |c|  }
\hline
~~~        & Au+Au $\snn$ = 200 GeV        & Pb+Pb $\snn$ = 2.76 TeV         \\
\hline
$\tau_{0}$ (fm)             & 0.6        & 0.6              \\
\hline
 $K$ (GeV/fm$^{3}$ )                   & 35.5       & 103.0            \\
\hline
$\alpha$                      & 0.05       & 0.05        \\
\hline
$\sigma_\text{NN} \text{(mb)}$                 & 42       & 64        \\
\hline
$\eta_{w}$                    & 1.3      & 2.0        \\
\hline
$\sigma_{\eta}$               & 1.5~      & 1.8       \\
\hline
\end{tabular}
\caption{\label{t:modelparameters} Common parameters shared between different initial condition parametrizations~\cite{Pang:2018zzo,Loizides:2017ack}.}
\end{center}
\end{table}

The initial fluid velocity at $\tau_{0}$ is assumed to follow the Bjorken approximation in this work as $v_{x} = v_{y} =0$ and $v_{z} = z/t$, where the initial transverse flow and the asymmetric distribution of $v_z$ along the $x$ direction are ignored. More sophisticated initial velocity profiles will be investigated in an upcoming study.

\subsection{Initial energy density, eccentricity and pressure gradient}

\begin{figure}[!tbp]
\begin{center}
\includegraphics[width=0.75\linewidth]{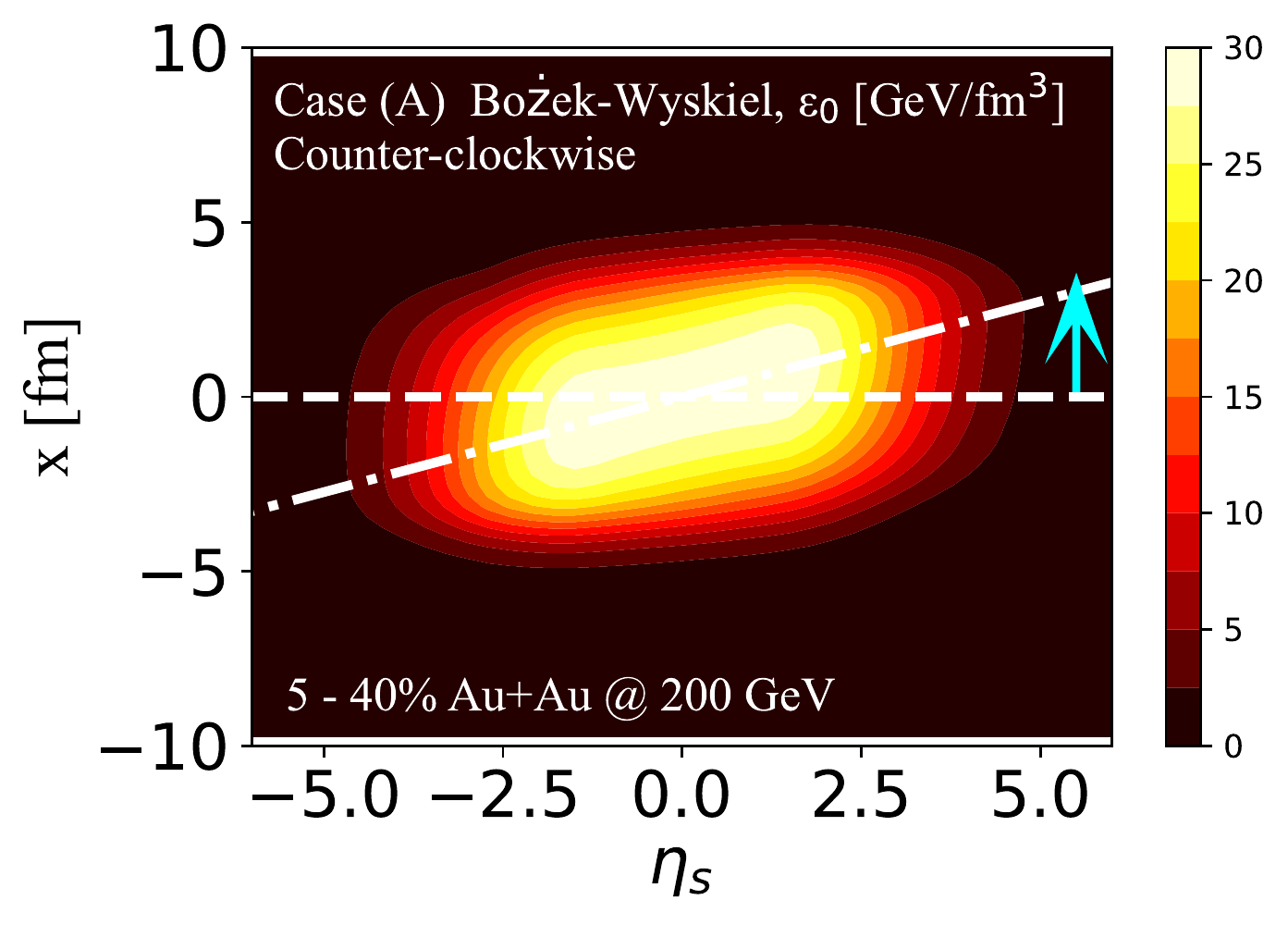}~\\
\includegraphics[width=0.75\linewidth]{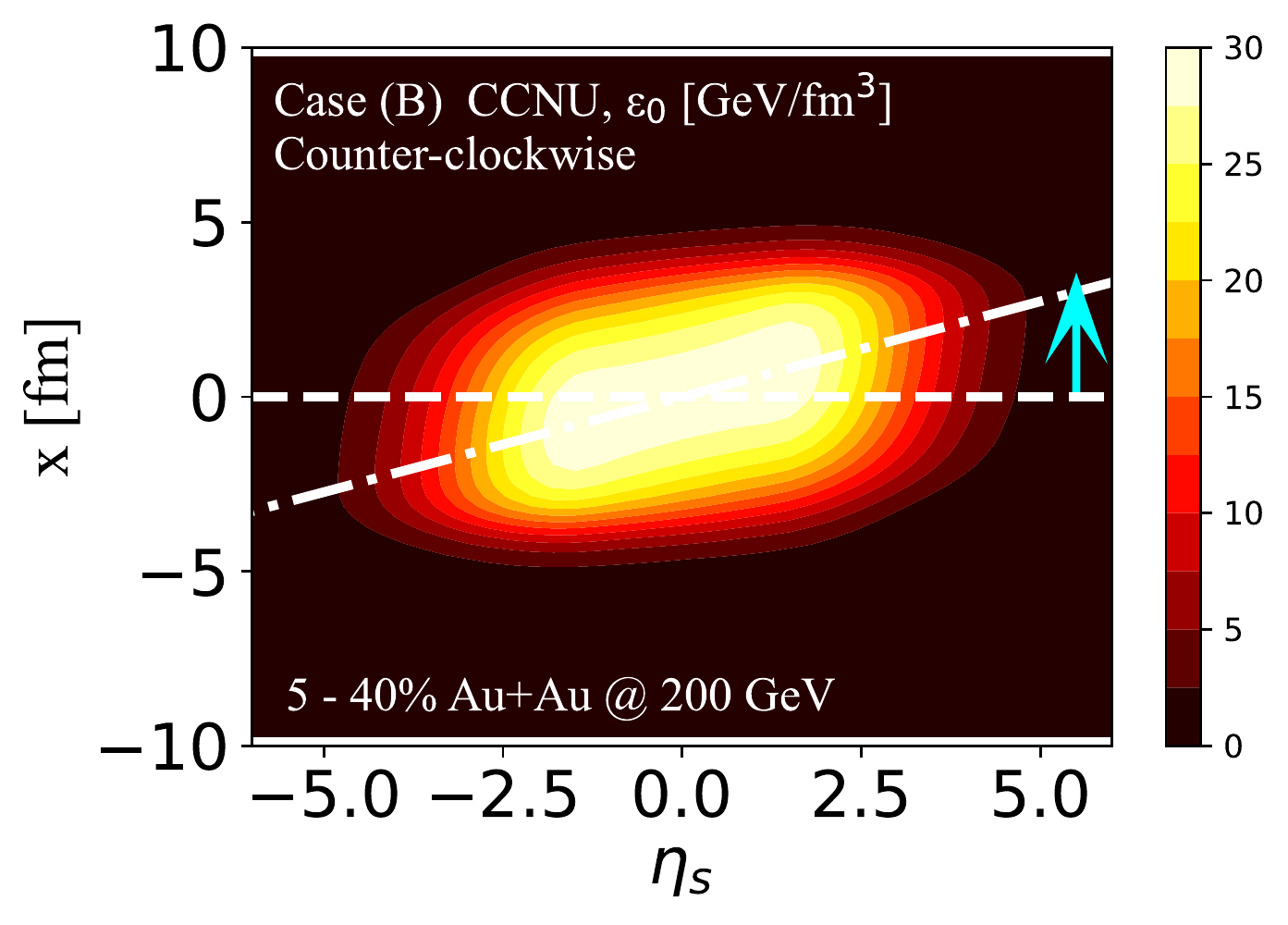}~\\
\includegraphics[width=0.75\linewidth]{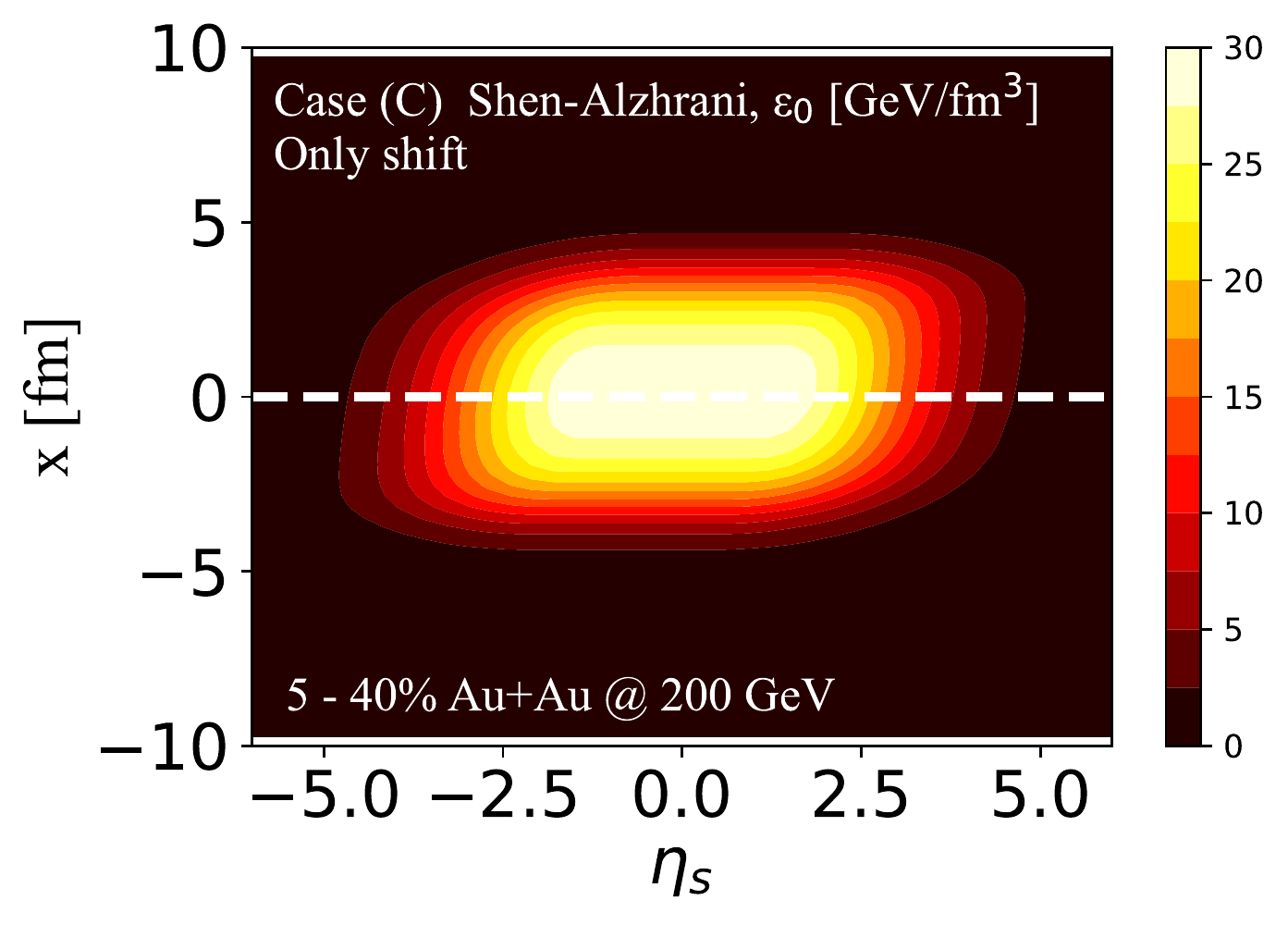}~
\end{center}
\caption{(Color online) The initial energy density on the $\eta_{s}$-$x$ plane at $\tau_0$ = 0.6~fm in 5-40\% ($b$ = 6.7~fm) Au+Au collisions at $\sqrt{s_\text{NN}}=200$~GeV. From top to bottom panel, we present Case (A) Boz$\dot{\textrm{e}}$k-Wyskiel parametrization with $\eta_{m} = 2.8$, Case (B) CCNU parametrization with $H_{t} = 2.9$ and Case (C) Shen-Alzhrani parametrization with $f$ = 0.15. The arrows (aqua color) sketch the counter-clockwise tilted initial condition with respect to the $x=0$ axis in the $\eta_{s}$-$x$ plane. }
\label{f:auau200ed}
\end{figure}

With the parametrizations above, we first compare the energy density profile between different setups.
In Fig.~\ref{f:auau200ed}, we present the initial energy density at $\tau_0$ = 0.6~fm on the $\eta_{s} - x$ plane for 5-40\% Au+Au collisions at $\snn = 200$~GeV. Three different parmetrizations of the initial energy distribution are compared. In order to describe the directed flow of charged particles later, the model parameter $\eta_{m} = 2.8$ is taken for the Boz$\dot{\textrm{e}}$k-Wyskiel parametrization (top panel) and $H_{t} = 2.9$ for the CCNU parametrization (middle panel), while $f = 0.15$ is taken from Ref.~\cite{Ryu:2021lnx} for Shen-Alzhrani parametrization (bottom panel). From Fig.~\ref{f:auau200ed}, we observe that parametrizations in our Case (A) and (B) generate similar initial energy density profiles: the distribution is not only shifted in the forward/backward rapidity direction for the positive/negative $x$ region, it is also tilted counter-clockwise relative to $x=0$ in the $\eta_{s}$-$x$ plane (following the arrow direction in the figure). On the other hand, the distribution from Case (C) only appears shifted horizontally in the rapidity direction. We note that our Case (C) here is similar to the Hirano-Tsuda parametrization~\cite{Hirano:2002ds,Hirano:2005xf} as illustrated in Ref.~\cite{Bozek:2010bi}.

\begin{figure}[!tbp]
\begin{center}
\includegraphics[width=0.75\linewidth]{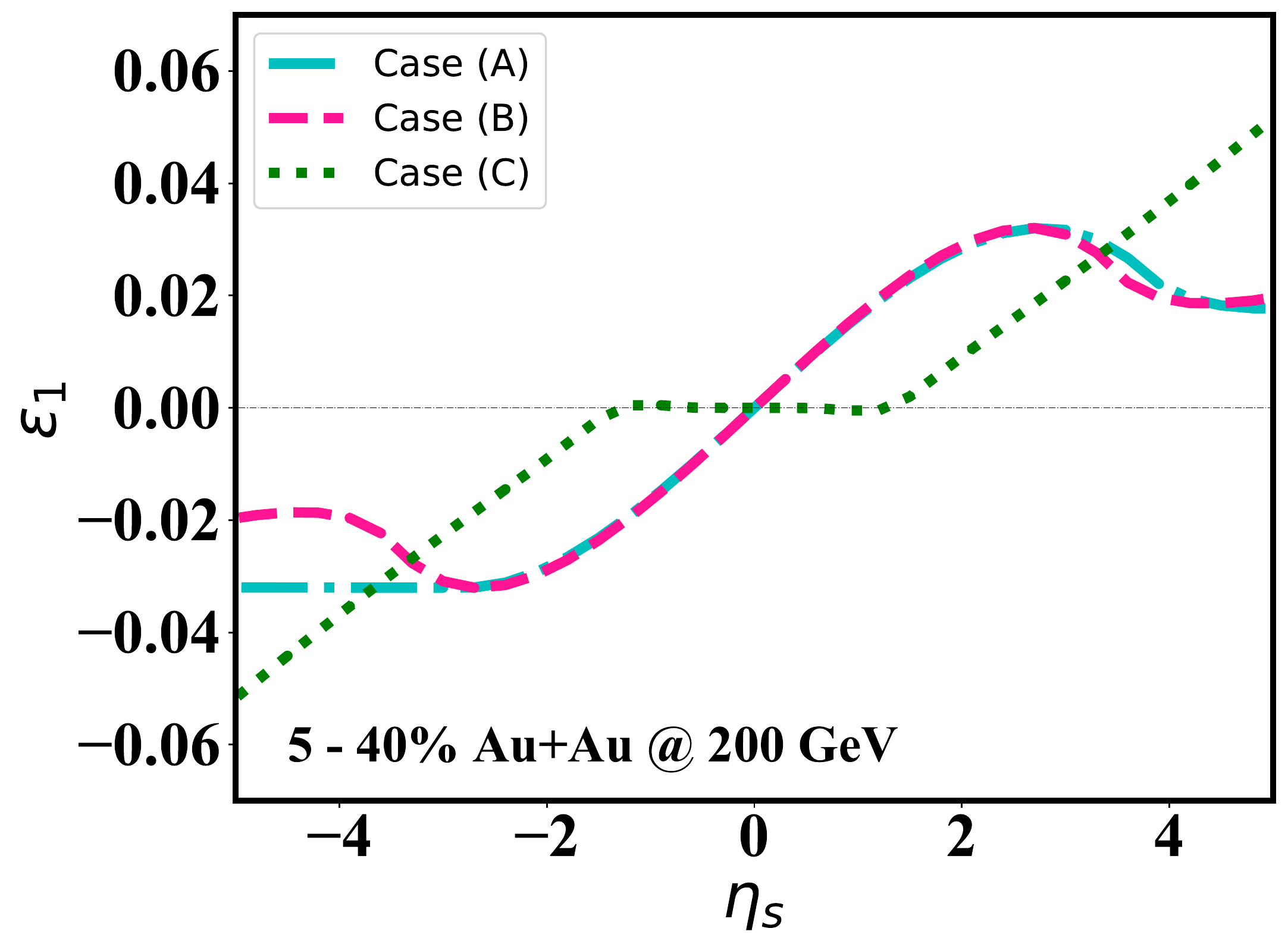}
\end{center}
\caption{(Color online) The first-order eccentricity coefficient $\varepsilon_{1}(\eta_{s})$ at $\tau_0$ = 0.6~fm for 5-40\% Au+Au collisions at $\sqrt{s_\text{NN}}=200$ GeV, compared between three different model setups.}
\label{f:auau200ecc1}
\end{figure}

In order to quantify the asymmetry of the initial energy density distribution with respect to the $y$-$z$ plane from different models, we present their corresponding first-order eccentricity coefficient $\varepsilon_1$ in Fig.~\ref{f:auau200ecc1} as a function of the space-time rapidity. The first-order eccentricity vector is defined as~\cite{Qiu:2011iv,Shen:2020jwv}:
\begin{equation}
\begin{aligned}
\vec{\mathcal{E}}_{1}\equiv\varepsilon_{1}(\eta_{s})e^{i\Psi_{1}(\eta_{s})}=
- \frac{\int d^{2}r \widetilde{r}^{3} e^{i\widetilde{\phi}}\varepsilon(r,\phi,\eta_{s})}{\int d^{2}r \widetilde{r}^{3} \varepsilon(r,\phi,\eta_{s})},
\label{eq:ecc1}
\end{aligned}
\end{equation}
in which the angular distribution is evaluated with respect to the  center-of-mass $(x_{0}(\eta_{s}), y_{0}(\eta_{s}))$ of each rapidity slice given by
\begin{equation}
\begin{aligned}
x_{0}(\eta_{s}) = \frac{\int d^{2} r x \varepsilon(r,\phi,\eta_{s})}{\int d^{2}r \varepsilon(r,\phi,\eta_{s})},
\label{eq:xy1}
\end{aligned}
\end{equation}
\begin{equation}
\begin{aligned}
y_{0}(\eta_{s}) = \frac{\int d^{2} r y \varepsilon(r,\phi,\eta_{s})}{\int d^{2}r \varepsilon(r,\phi,\eta_{s})}.
\end{aligned}
\end{equation}
The transverse radius and the azimuthal angle are then defined as $\widetilde{r}(x,y,\eta_{s})=\sqrt{(x-x_{0})^{2}+(y-y_{0})^{2}}$ and $\widetilde{\phi}(x,y,\eta_{s})=\arctan[(y-y_{0})/(x-x_{0})]$ respectively. In the end, $\varepsilon_1$ in Eq.~(\ref{eq:ecc1}) gives the first-order eccentricity coefficient while $\Psi_1$ gives the corresponding participant plane angle. The $\eta_s$ dependence of this $\vec{\mathcal{E}}_{1}$ will contribute to explaining what kind of longitudinally deformed fireball is needed to produce the final hadron $v_{1}$.

In Fig.~\ref{f:auau200ecc1}, one observes similar $\eta_s$ dependence of $\varepsilon_1$ between Case (A) and (B), they are odd functions of $\eta_s$ and positive/negative in the $+$/$-\eta_s$ regime.
On the other hand, although $\varepsilon_1$ from Case (C) is also an odd function of $\eta_s$, its value is much smaller than that from (A) and (B) within $|\eta_s| < 2$. This will further affect the evolution profile of the nuclear medium in the subsequent hydrodynamic expansion.


\begin{figure}[!tbp]
\begin{center}
\includegraphics[width=0.75\linewidth]{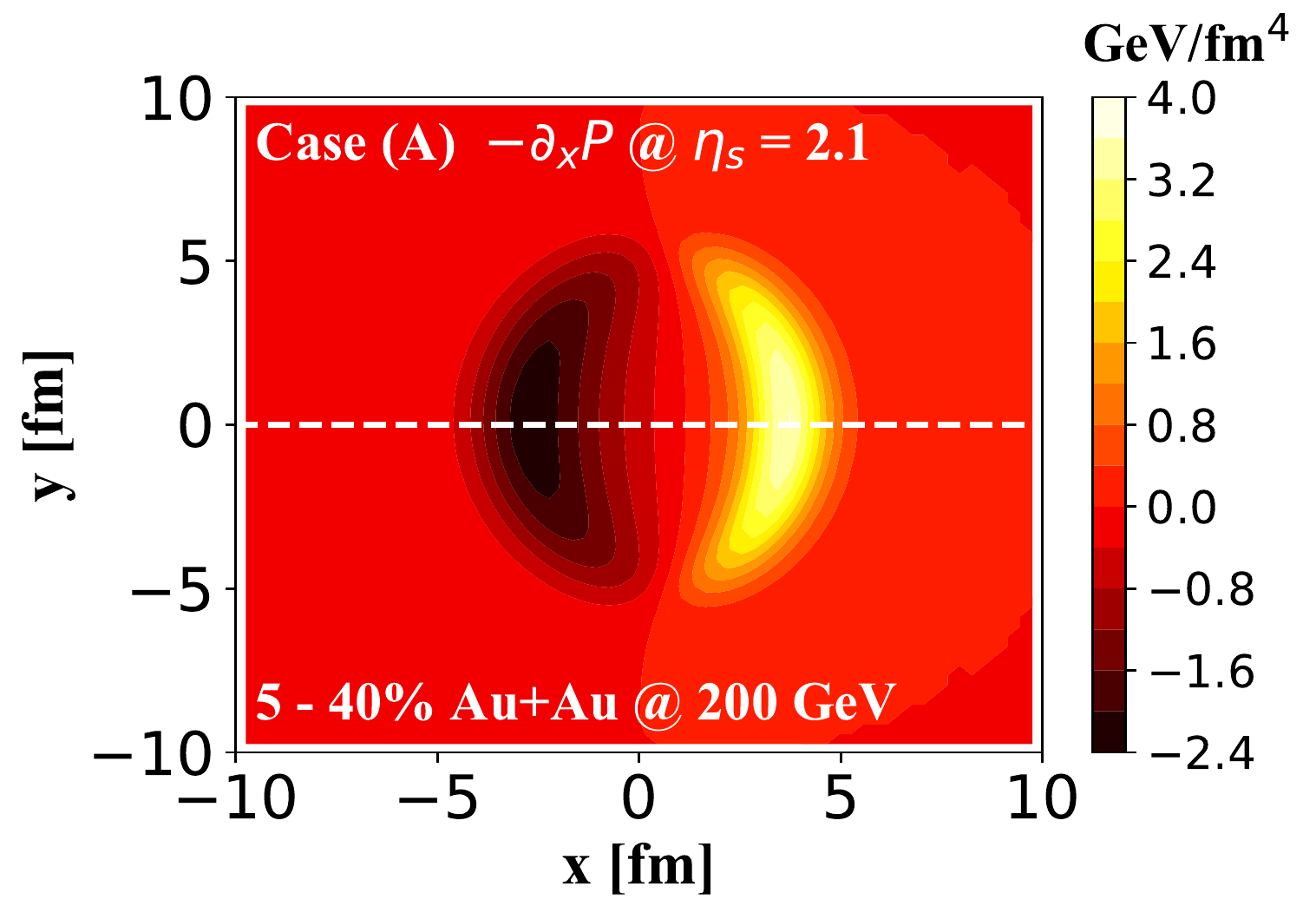}~\\
\includegraphics[width=0.75\linewidth]{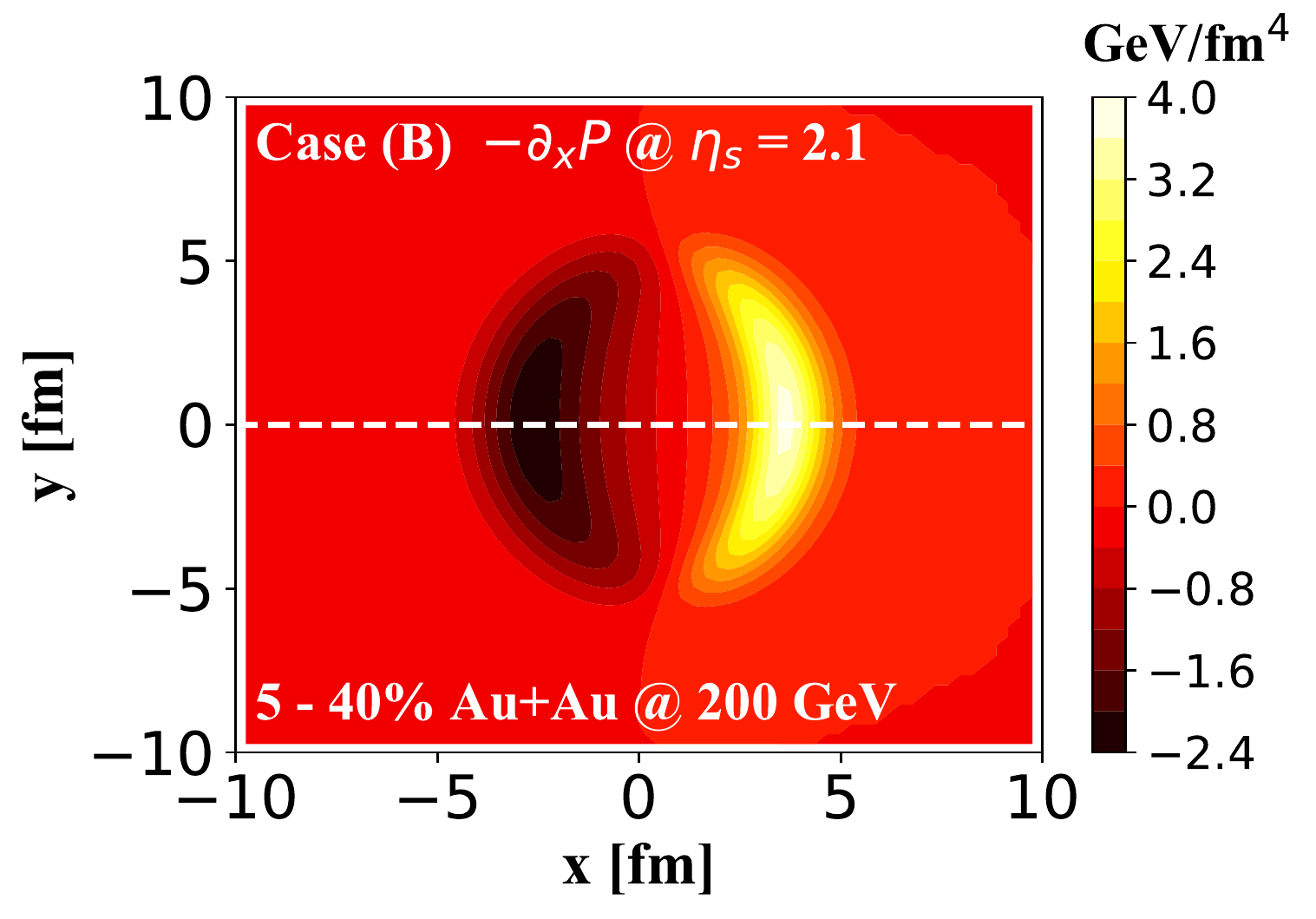}~\\
\includegraphics[width=0.75\linewidth]{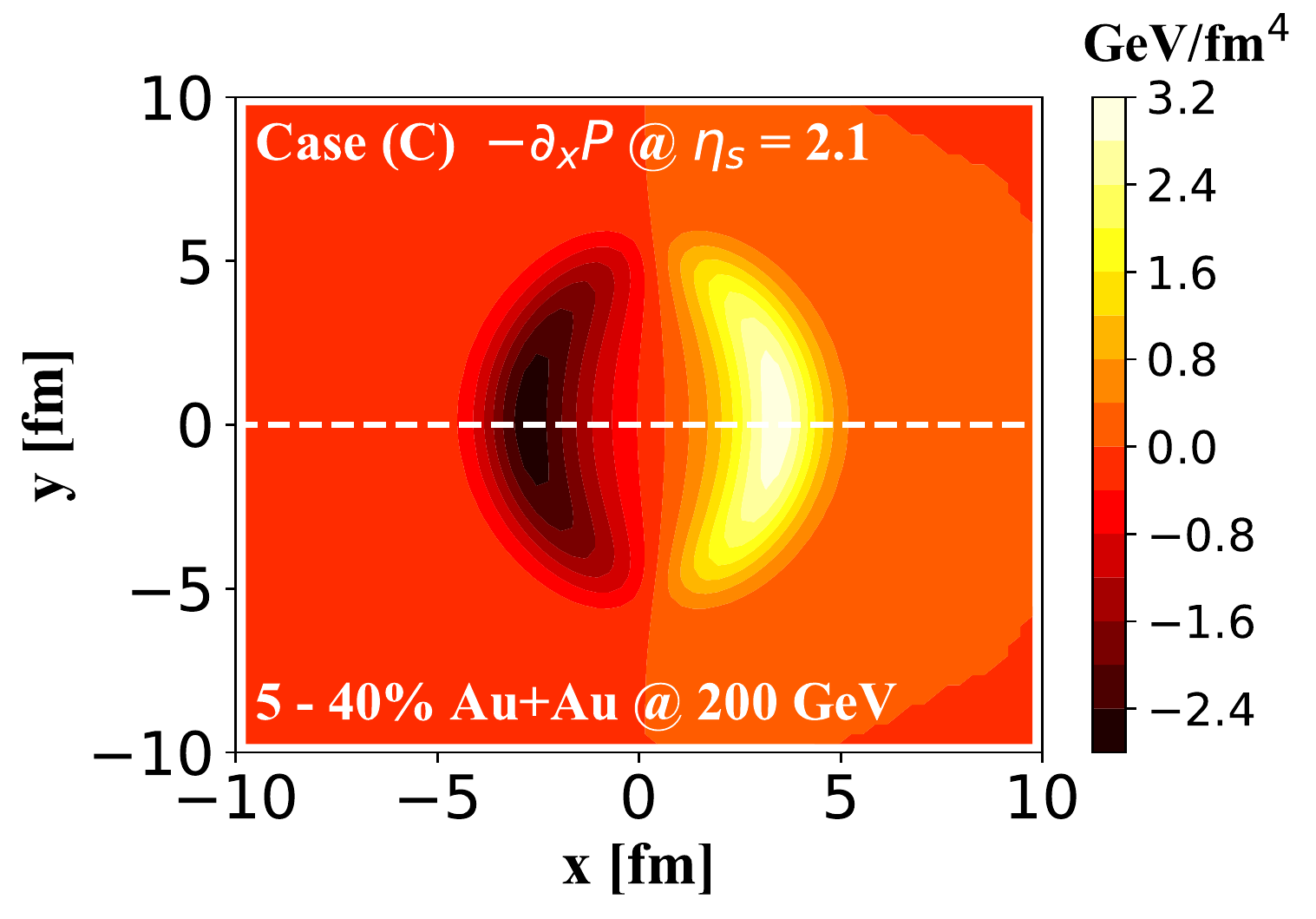}~
\end{center}
\caption{(Color online) The initial pressure gradient $-\partial_{x}P$ on the $x$-$y$ plane at $\tau_0 = 0.6$~fm and $\eta_s=2.1$ in 5-40\% ($b$ = 6.7~fm) Au+Au collisions at $\sqrt{s_\text{NN}}=200$~GeV, compared between three different model setups.}
\label{f:auau200pg}
\end{figure}

In addition to the energy density distribution, we also present the initial pressure gradient $-\partial_{x}P$ in the transverse plane, which will directly drive the development of radial flow of nuclear matter. Our three model setups are compared in Fig.~\ref{f:auau200pg} for the initial $-\partial_{x}P$ distribution in the $x$-$y$ plane at a given $\eta_s$, where the same parameter values of $\eta_{m}$, $H_{t}$, $f$ are used as for Fig.~\ref{f:auau200ed}. One may clearly observe the positive/negative value of $-\partial_{x}P$ in the $+$/$-x$ direction that drives the outward expansion of the medium. In the top and middle panels, we see that at forward rapidity, the center (zero pressure) regions of these distributions are shifted towards $+x$ for Case (A) and (B) due to the counter-clockwise tilt of the energy density distribution as previously discussed in Fig.~\ref{f:auau200ed}. To the contrary, such shift is weaker in the bottom panel here for Case (C) due to its different profile of energy density. Whether the average $x$-component of the final-state hadron momentum will be positive or negative at a given rapidity depends on the average value of this $-\partial_{x}P$ over the corresponding transverse plane and how it evolves with time. This will be discussed in detail soon in this work.


\subsection{Hydrodynamic evolution}
Starting with the initial energy density and flow velocity described above, we use the (3+1)-D viscous hydrodynamic model CLVisc~\cite{Pang:2016igs,Pang:2018zzo,Wu:2018cpc,Chen:2017zte,He:2018gks} to simulate the subsequent evolution of the QGP medium in this work. The hydrodynamic equation reads~\cite{Jiang:2020big,Jiang:2018qxd,Denicol:2012cn,Romatschke:2009im,Romatschke:2017ejr}
\begin{equation}
\begin{aligned}
\partial_{\mu}T^{\mu\nu}=0,
\label{eq:tmn}
\end{aligned}
\end{equation}
where $T^{\mu\nu}$ is the energy-momentum tensor defined as
\begin{equation}
\begin{aligned}
T^{\mu\nu}=\varepsilon u^{\mu}u^{\nu}-(P+\Pi)\Delta^{\mu\nu} + \pi^{\mu\nu}.
\label{eq:tensor}
\end{aligned}
\end{equation}
It is composed of the local energy density $\varepsilon$,
the fluid four-velocity $u^{\mu}$, the pressure $P$, the shear stress tensor $\pi^{\mu\nu}$ and the bulk pressure $\Pi$.
The projection tensor is given by $\Delta^{\mu\nu} = g^{\mu\nu}-u^{\mu}u^{\nu}$, and the metric tensor $g^{\mu\nu} = \text{diag} (1,-1,-1,-1)$ is used. The hydrodynamic equations are solved together with the lattice QCD Equation of State (EoS) from the Wuppertal-Budapest group (2014)~\cite{Borsanyi:2013bia}, and the shear-viscosity-to-entropy-density ratio is set as $\eta_{v}/s = 0.08$ ($\eta_{v}$ for the shear viscosity) for all collision systems investigated in this work. However, we have ignored effects of bulk viscosity and net baryon density at this moment, which have been incorporated in the recent CLVisc development~\cite{Zhao:2021vmu,Wu:2021fjf} and will also be taken into account in our follow-up effort.

The isothermal freeze-out condition~\cite{Pang:2018zzo} is applied in the current calculation, where the freeze-out hypersurface is determined by a constant temperature value $T_\text{frz}=137$~MeV. On this hypersurface, hadron spectra are evaluated based on the Cooper-Frye formalism~\cite{Cooper:1974mv}. Contributions from resonance decay have also been taken into account according to Ref.~\cite{Pang:2018zzo}.

\subsection{Time evolution of average pressure gradient and flow velocity}

\begin{figure}[!tbp]
\begin{center}
\includegraphics[width=0.8\linewidth]{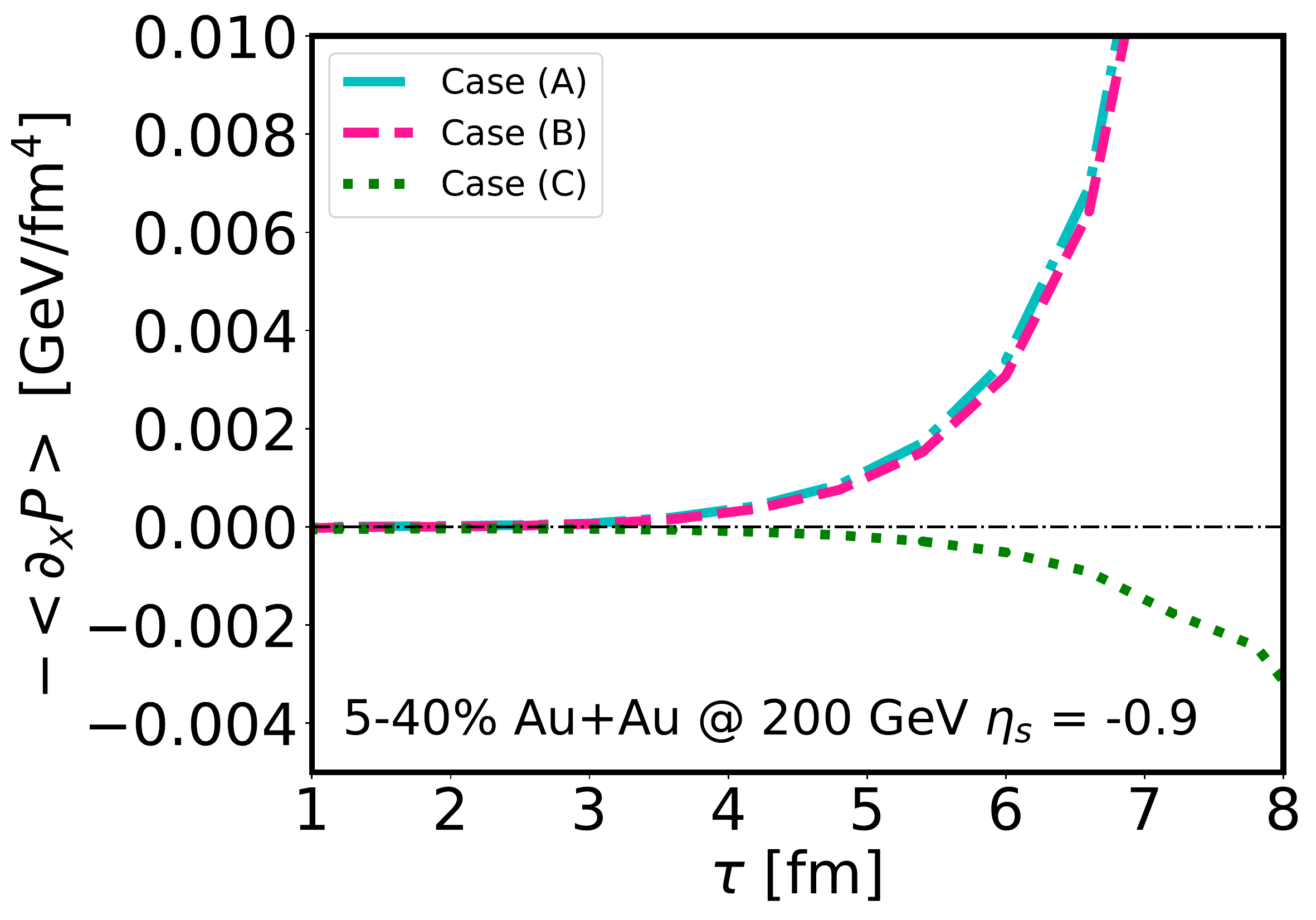}
\includegraphics[width=0.8\linewidth]{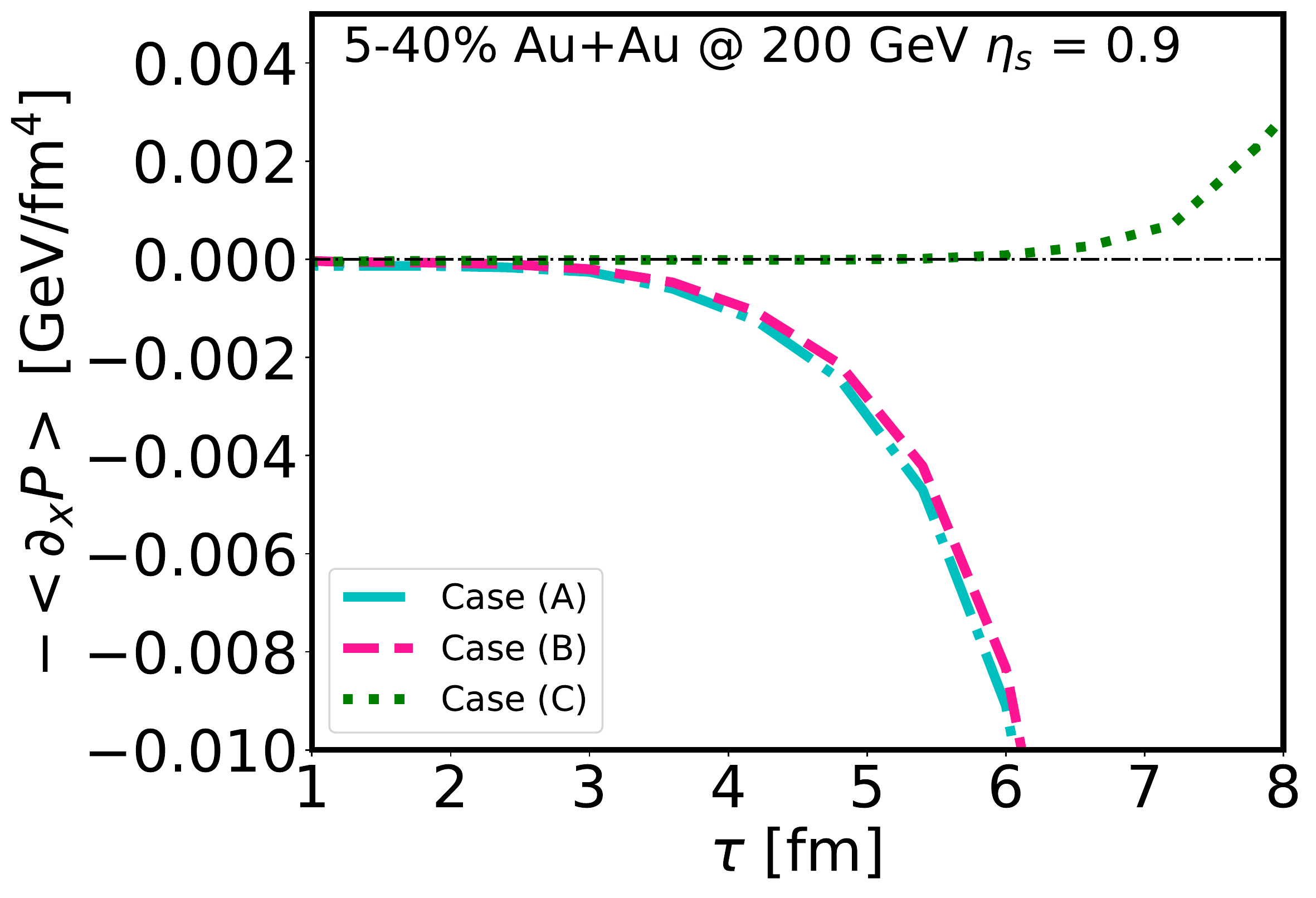}
\end{center}
\caption{(Color online) Time evolution of the average pressure gradient in the $x$ direction at $\eta_{s} = -0.9$ (upper panel) and $\eta_{s} = 0.9$ (lower panel) in 5-40\% Au+Au collisions at $\sqrt{s_\text{NN}}=200$~GeV, compared between three parametrization methods of the initial condition.}
\label{f:auau200pge1}
\end{figure}

\begin{figure}[!tbp]
\begin{center}
\includegraphics[width=0.8\linewidth]{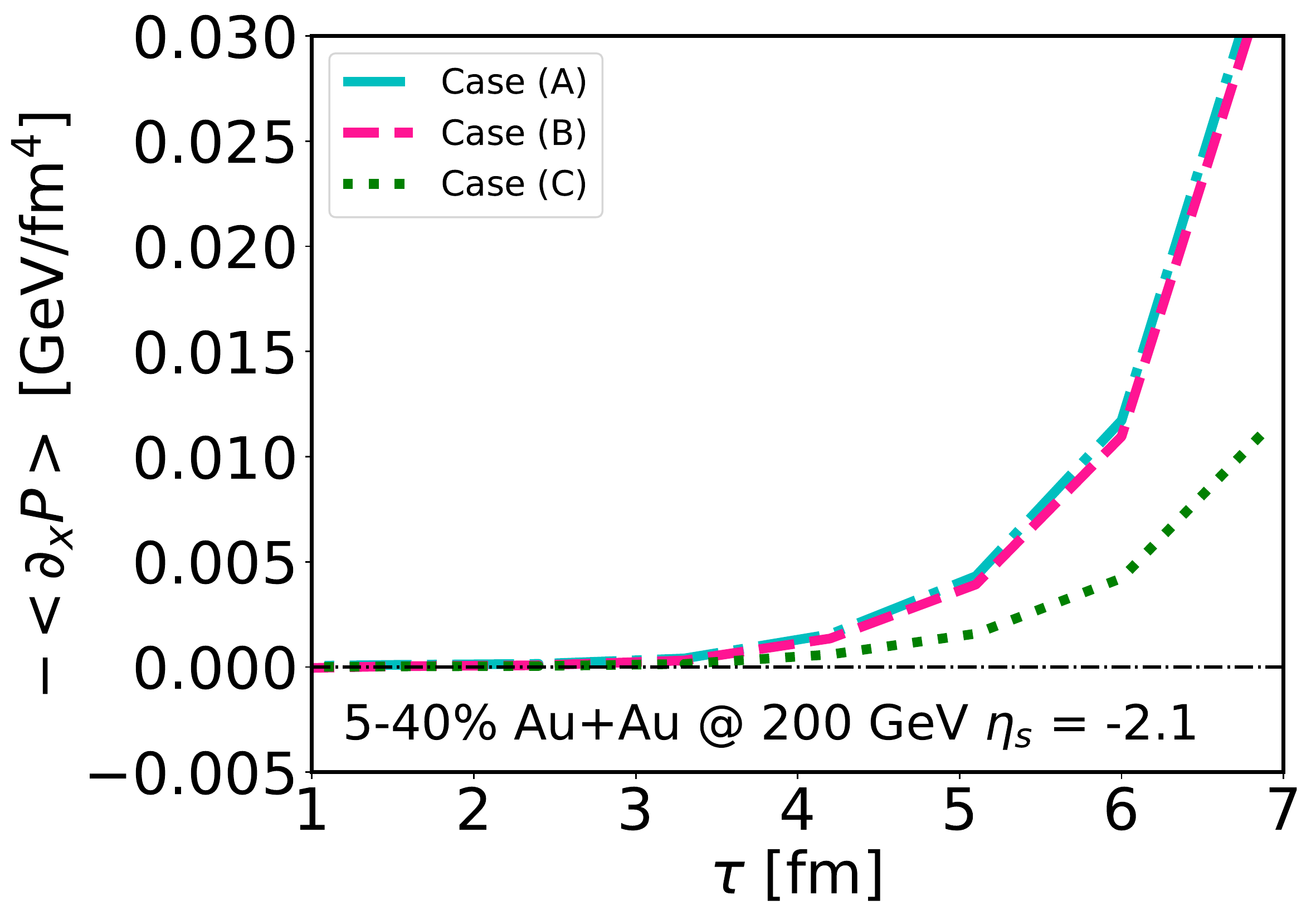}
\includegraphics[width=0.8\linewidth]{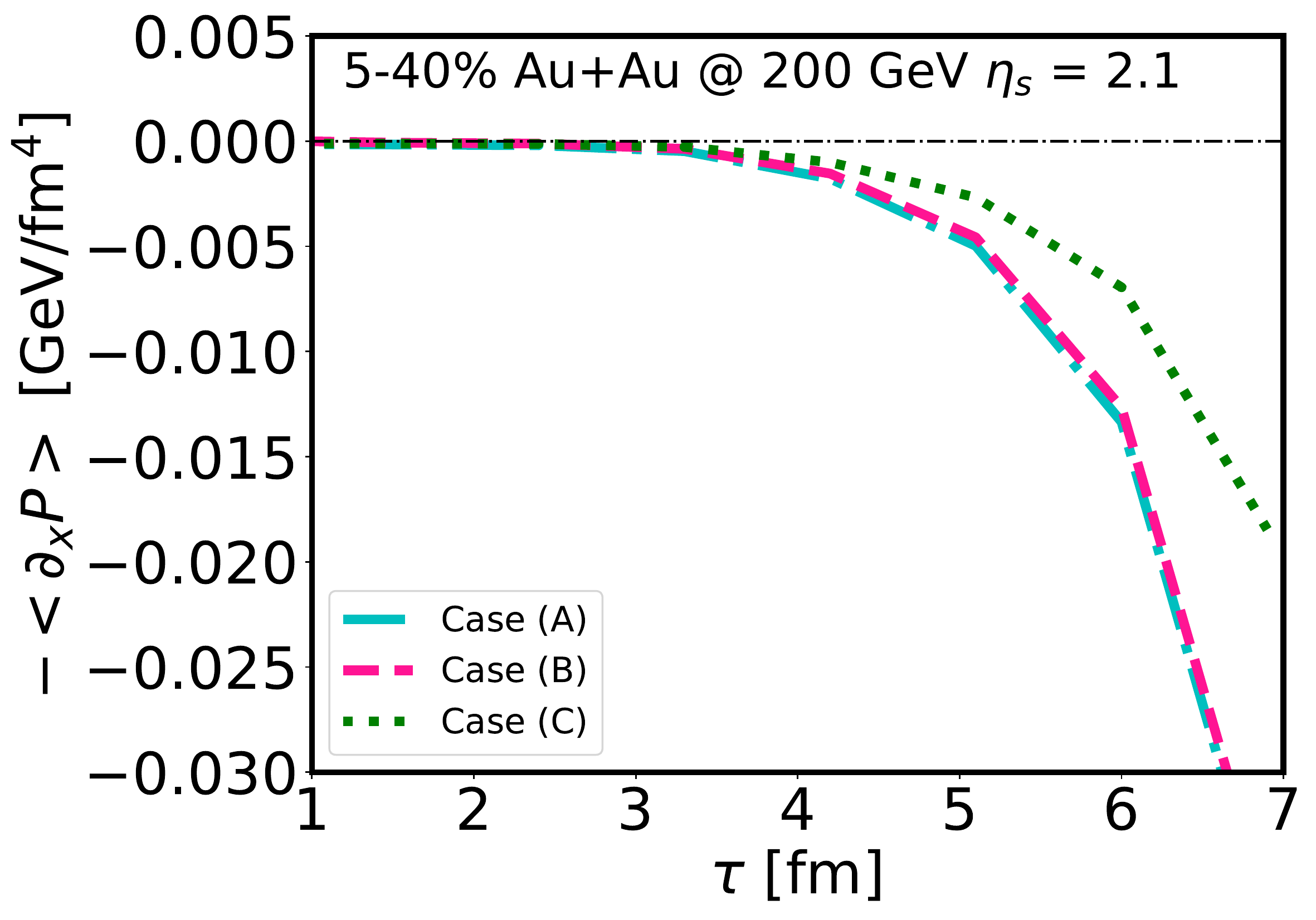}
\end{center}
\caption{(Color online) Time evolution of the average pressure gradient in the $x$ direction at $\eta_{s} = -2.1$ (upper panel) and $\eta_{s} = 2.1$ (lower panel) in 5-40\% Au+Au collisions at $\sqrt{s_\text{NN}}=200$~GeV, compared between three parametrization methods of the initial condition.}
\label{f:auau200pge2}
\end{figure}

Hydrodynamic model describes how the asymmetry of the initial energy density distribution is transferred to the anisotropy of the final-state hadron momentum. In this subsection, using the hyrodynamic simulation, we investigate how the average pressure gradient $-\langle\partial_{x}P \rangle$ and flow velocity $\langle v_{x} \rangle$ develop with time at different $\eta_s$. This will help understand the origin of directed flow and how it depends on the initial geometry of the nuclear matter.

As previously shown in Fig.~\ref{f:auau200pg}, the deformation of the initial energy density breaks the symmetry of the pressure gradient along the $x$ direction. In Fig.~\ref{f:auau200pge1}, we study how the average pressure gradient over the transverse plane at a given space-time rapidity ($\pm 0.9$ here) evolves with time. Similar to before, nuclear matter produced in 5-40\% Au+Au collisions at $\snn = 200$~GeV is used.
One finds that the time evolution of  $-\langle\partial_{x}P \rangle$ is significantly affected by the initial condition of the medium. For Case (A) and (B) of our initial parametrization, $-\langle\partial_{x}P \rangle$ increases from zero with time at $\eta_s = -0.9$ while decreases at $\eta_s = 0.9$, indicating an increasing overall force that accelerates the medium expansion toward the $+x$ direction at $\eta_s = -0.9$ while toward the $-x$ direction at $\eta_s = 0.9$. Little difference is observed between the two parametrization methods in Case (A) and (B). To the contrary, Case (C) leads to a qualitatively opposite average pressure gradient, therefore accelerating force, which slightly decreases from zero at at $\eta_s = -0.9$ while increases at $\eta_s = 0.9$. Note that unlike higher-order components of anisotropy, values of eccentricity $\varepsilon_1$ (in Fig.~\ref{f:auau200ecc1}) and pressure gradient $-\langle\partial_{x}P \rangle$ are not necessarily positively correlated to each other. Due to the tilted deformation of medium profiles in Case (A) and (B), its expansion in space contributes to an overall force on the $+$/$-x$ direction at backward/forward rapidity, as intuitively illustrated in Ref.~\cite{Bozek:2010bi}. However, since the initial energy density given by Case (C) is rather symmetric along the $\pm x$ direction at small $|\eta_s|$, different $-\langle\partial_{x}P \rangle$ is obtained.

The average pressure gradient can quantitatively or even qualitatively change as $|\eta_s|$ increases. Shown in Fig.~\ref{f:auau200pge2} are the time evolution of $-\langle\partial_{x}P \rangle$ at $\eta_s=-2.1$ (upper panel) and 2.1 (lower panel). Compared to the smaller $|\eta_s|$ region (Fig.~\ref{f:auau200pge1}), one observes a larger magnitude of the pressure gradient for Case (A) and (B). Meanwhile, a sign flip is also found for Case (C) due to its weaker asymmetric density distribution at smaller space-time rapidity but stronger asymmetric distribution at larger space-time rapidity along $\pm x$ direction, as can be seen in the bottom panel of Fig.~\ref{f:auau200ed}.


\begin{figure}[!tbp]
\begin{center}
\includegraphics[width=0.75\linewidth]{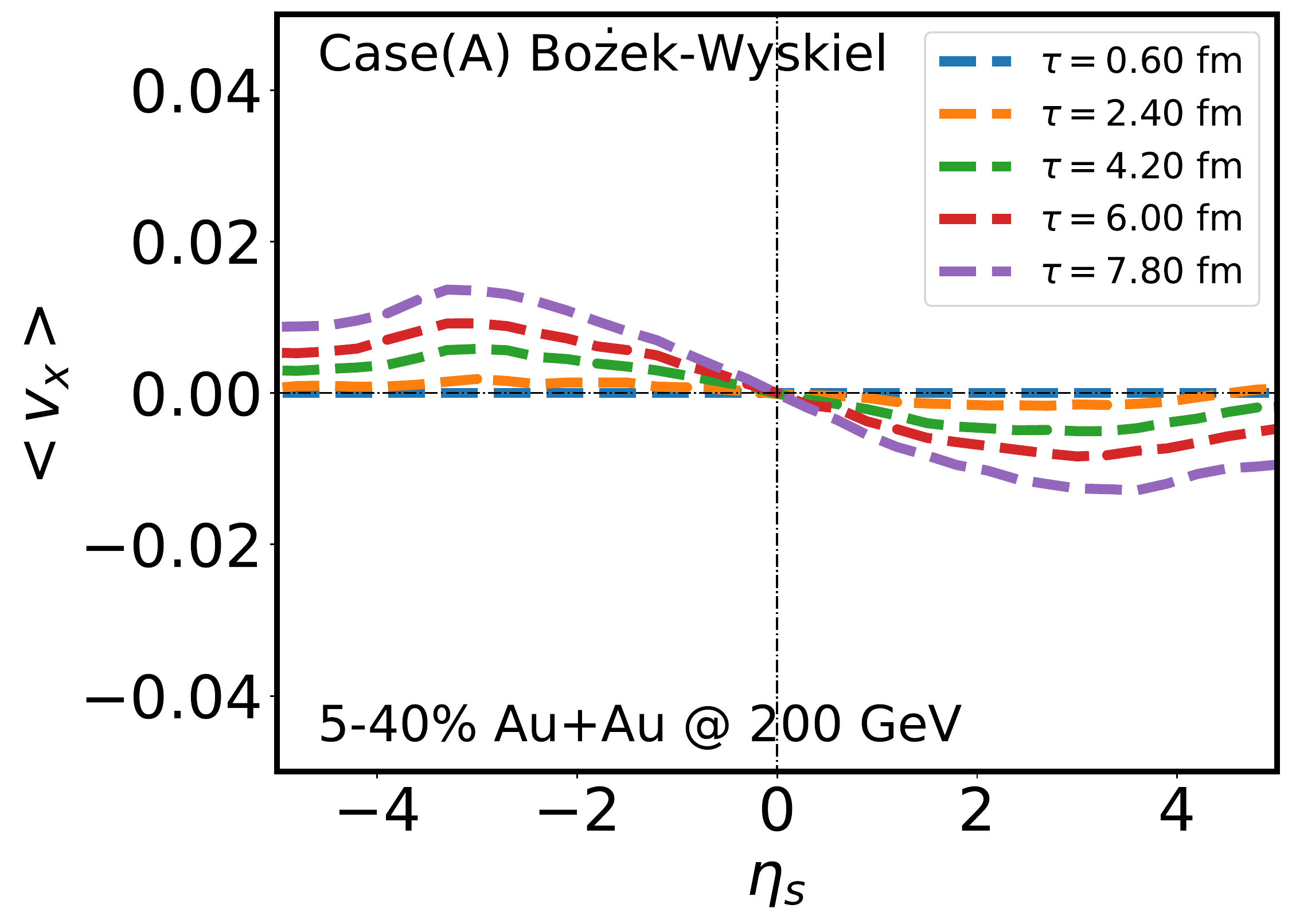}~\\
\includegraphics[width=0.75\linewidth]{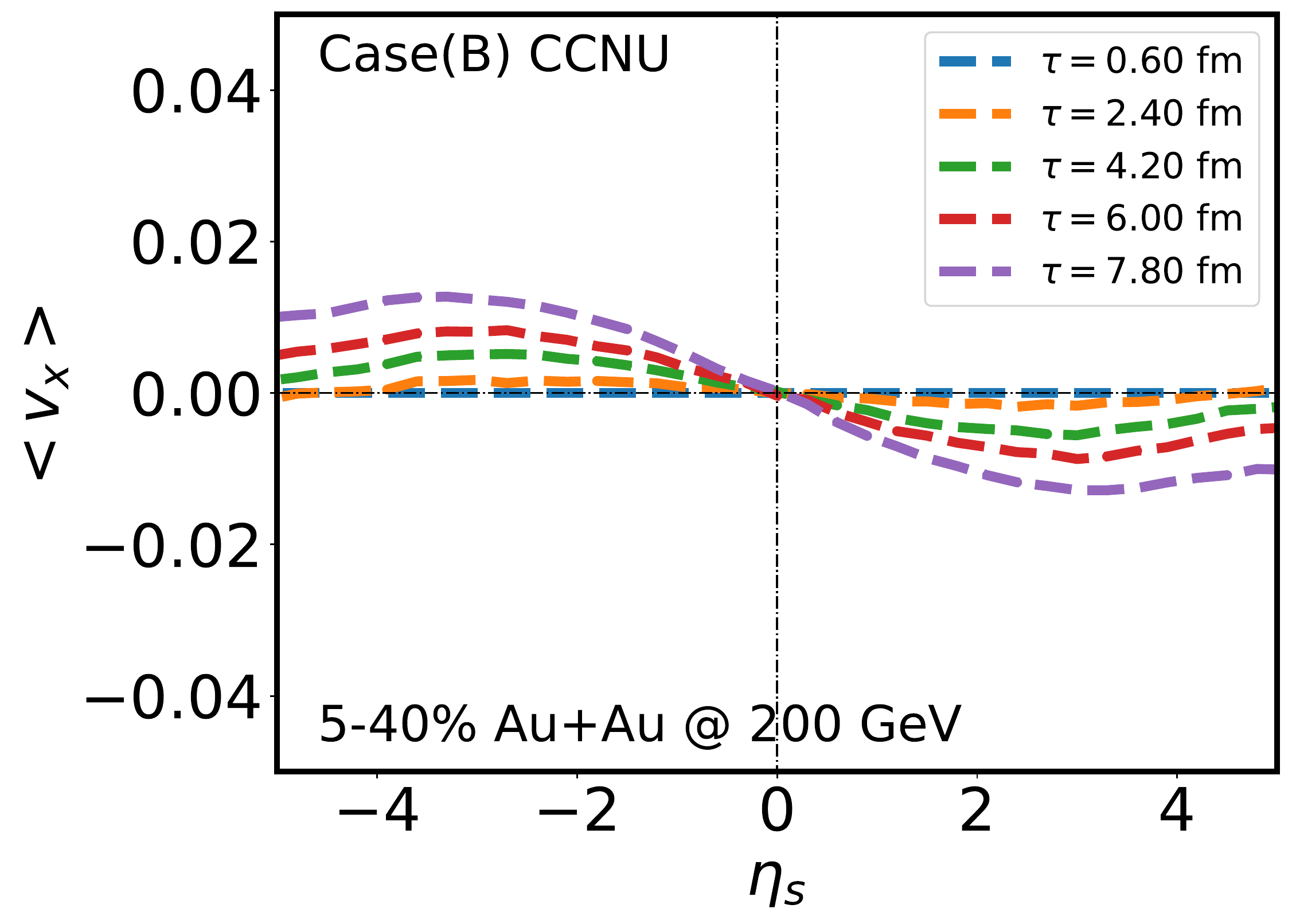}~\\
\includegraphics[width=0.75\linewidth]{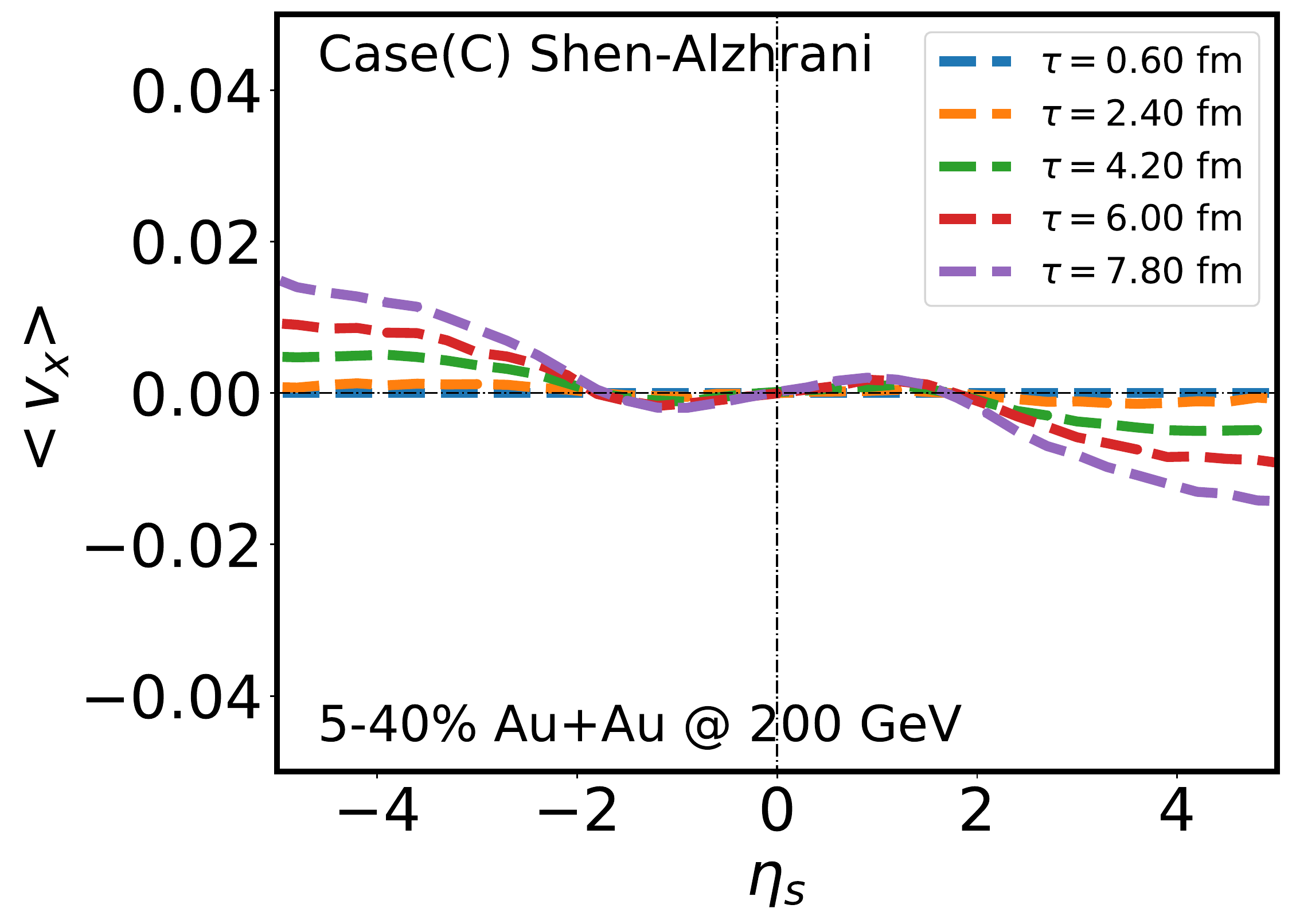}~
\end{center}
\caption{(Color online) Space-time rapidity dependence of the average flow velocity in the $x$ direction at different evolution times in 5-40\% Au+Au collisions at $\sqrt{s_\text{NN}}=200$~GeV, compared between three parametrization methods of the initial condition.}
\label{f:auau200vx}
\end{figure}

A direct outcome of different pressure gradients is the different flow velocities in the corresponding direction.
In Fig.~\ref{f:auau200vx}, we study how the space-time rapidity dependence of the average flow velocity $\langle v_{x} \rangle$ develops with time. The average flow velocity at a given time and rapidity can be evaluated as~\cite{Bozek:2010bi,Heinz:2013th},
\begin{equation}
\begin{aligned}
\langle v_{x}(\eta_{s}) \rangle = \frac{\int d^{2}r v_{x} \gamma \varepsilon(r,\phi,\eta_{s}) }{\int d^{2}r \gamma \varepsilon(r,\phi,\eta_{s})},
\label{eq:tar}
\end{aligned}
\end{equation}
where $\gamma=1/\sqrt{1-v_{x}^{2}-v_{y}^{2}-v_{\eta_{s}}^{2}}$ is the Lorentz boost factor.

Driven by the pressure gradient $-\langle\partial_{x}P \rangle$ previously shown in Figs.~\ref{f:auau200pge1} and~\ref{f:auau200pge2}, the average flow velocity $\langle v_x \rangle$ in Fig.~\ref{f:auau200vx} is positive/negative at backward/forward rapidity for Case (A) and (B). The magnitude of $\langle v_x \rangle$ increases with time as that of the pressure gradient does. A larger $|-\langle\partial_{x}P \rangle|$ generates a larger $\langle v_x \rangle$ at $|\eta_s|$ around $|\eta_s|\approx 2$ than around $|\eta_s|\approx 1$. In contrast, $\langle v_x \rangle$ from Case (C) shows opposite sign to that from Case (A) and (B) at small $|\eta_s|$, while the same sign beyond $|\eta_s|\approx 2$. This is also consistent with the behaviors of the average pressure gradient at different space-time rapidity from Case (C). The average flow velocity of the QGP medium here will directly contribute to the directed flow coefficient of the final-state hadrons emitted from the QGP. 


\section{Charged particle yield and directed flow}
\label{v1section3}

With the above setup of energy density initialization and hydrodynamic simulation, we present our results of charged particle yield and directed flow in this section. In particular, we investigate how the directed flow depends on the energy density distribution of the initial state.

\begin{figure}[!tbp]
\begin{center}
\includegraphics[width=0.75\linewidth]{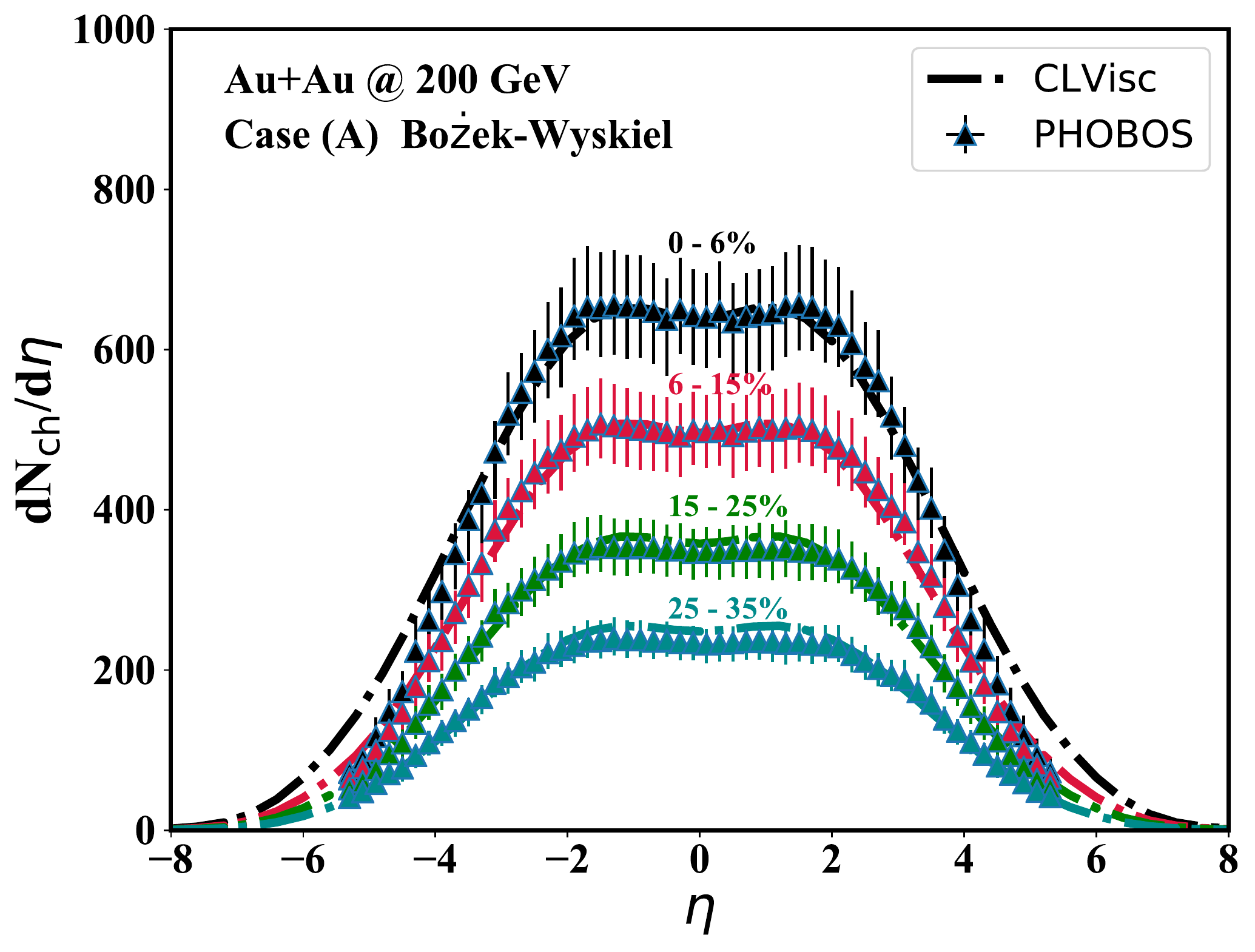}~\\
\includegraphics[width=0.75\linewidth]{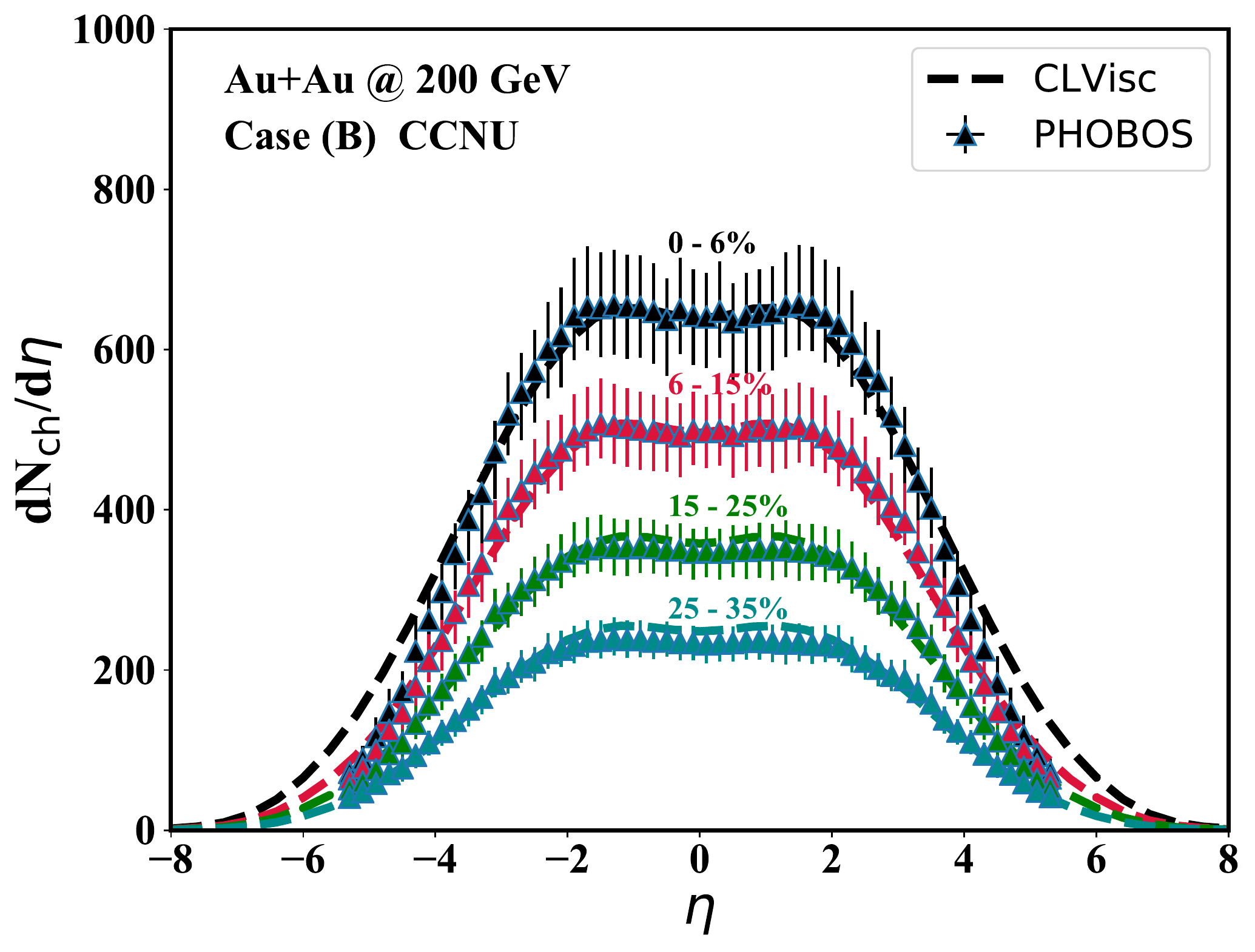}~\\
\includegraphics[width=0.75\linewidth]{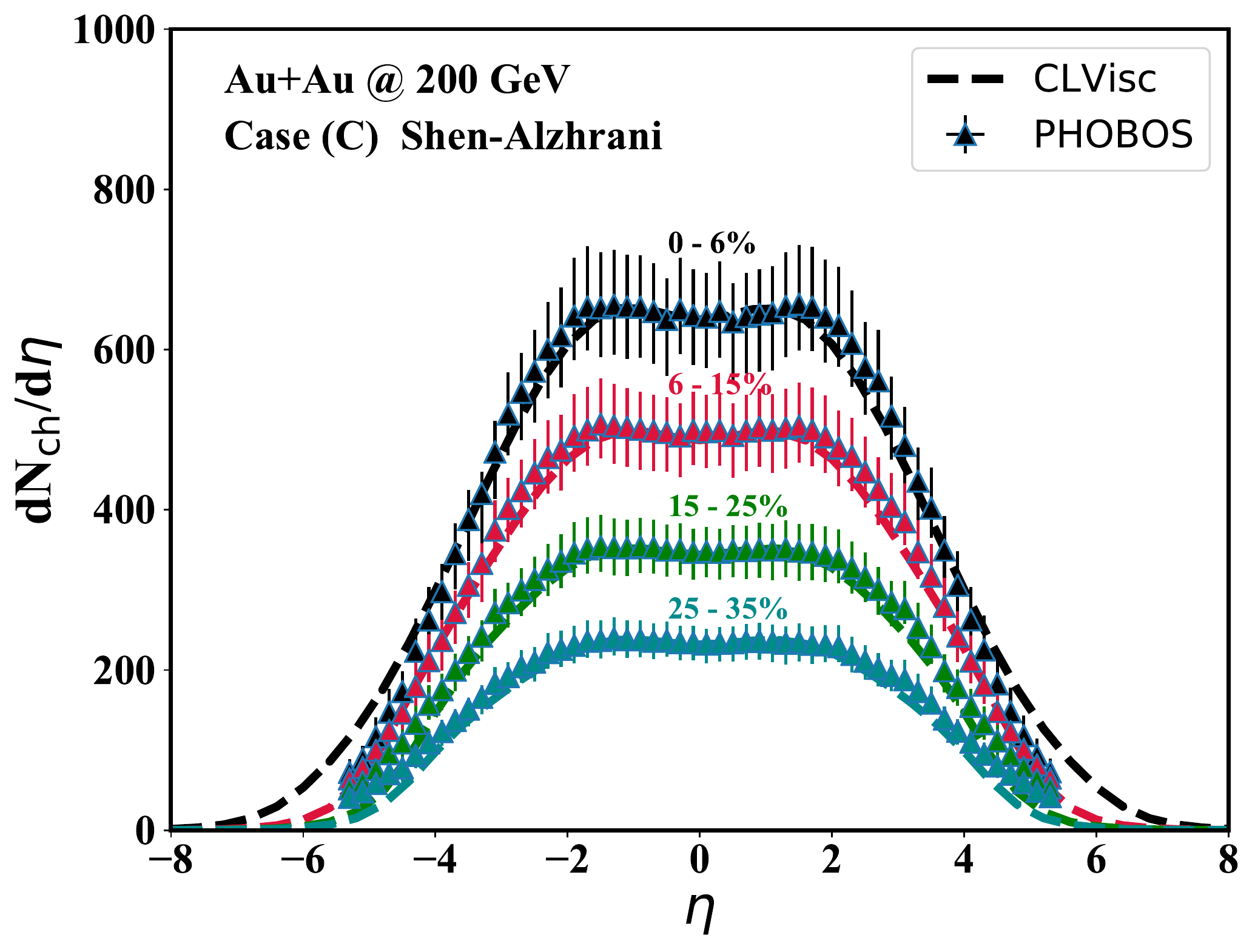}~
\end{center}
\caption{(Color online) Pseudorapidity distribution of charged particles in Au+Au collisions at $\snn = 200$~GeV in 0-6\%, 6-15\%, 15-25\% and 25-35\% centrality classes, compared between the CLVisc hydrodynamic calculation with three initial condition setups and the PHOBOS data~\cite{Alver:2010ck}.}
\label{f:auau200dndeta}
\end{figure}

\begin{figure}[!tbp]
\begin{center}
\includegraphics[width=0.75\linewidth]{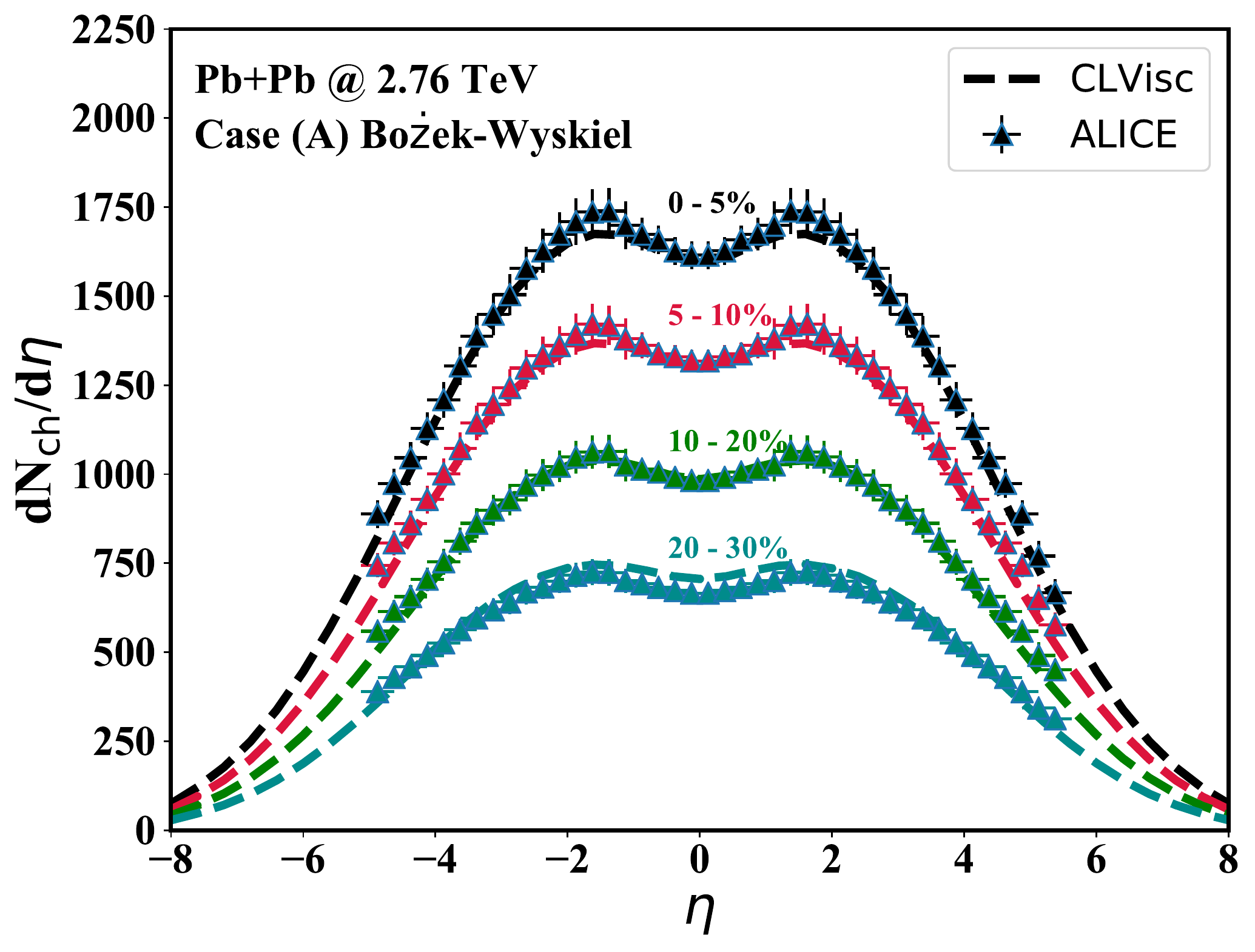}~\\
\includegraphics[width=0.75\linewidth]{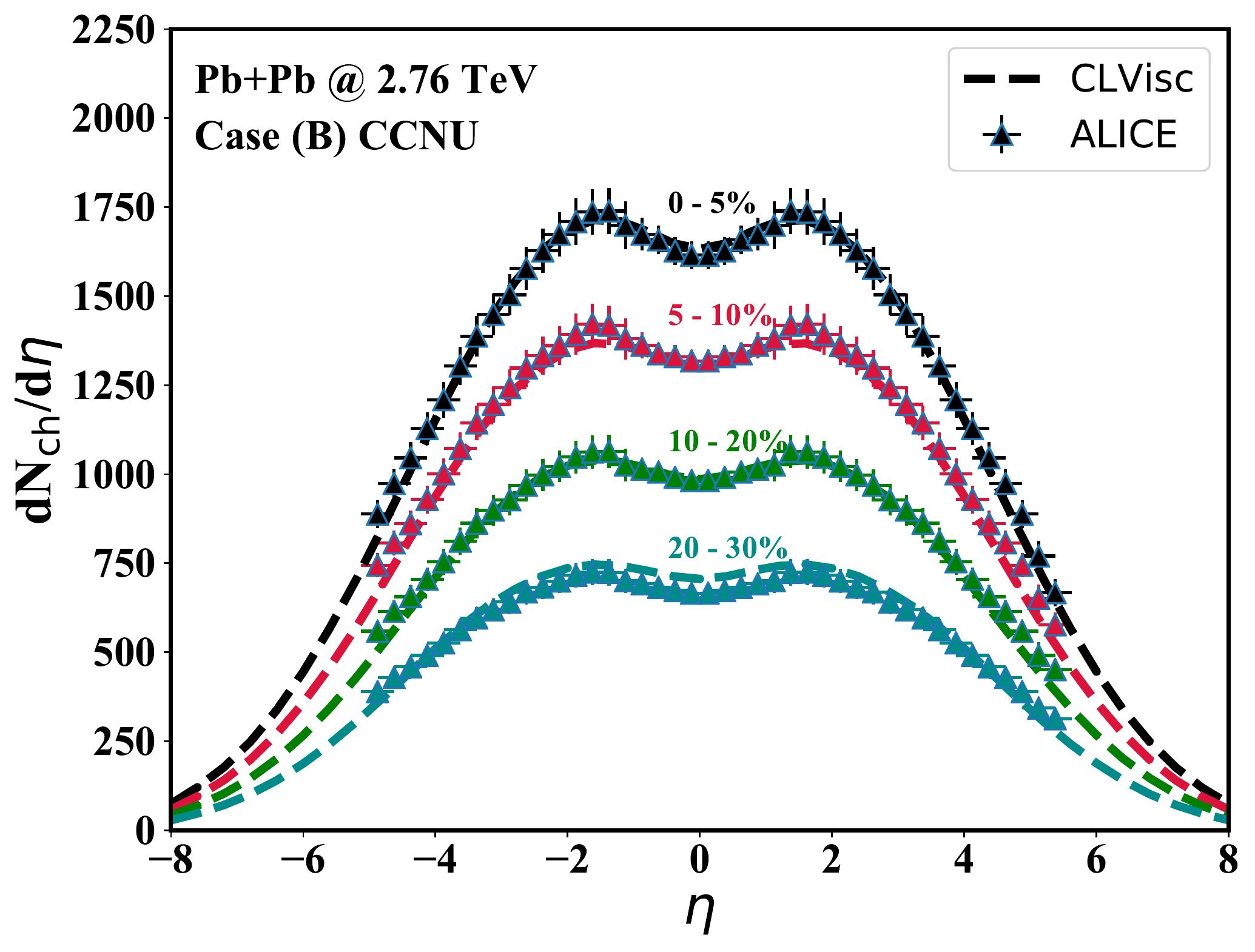}~\\
\includegraphics[width=0.75\linewidth]{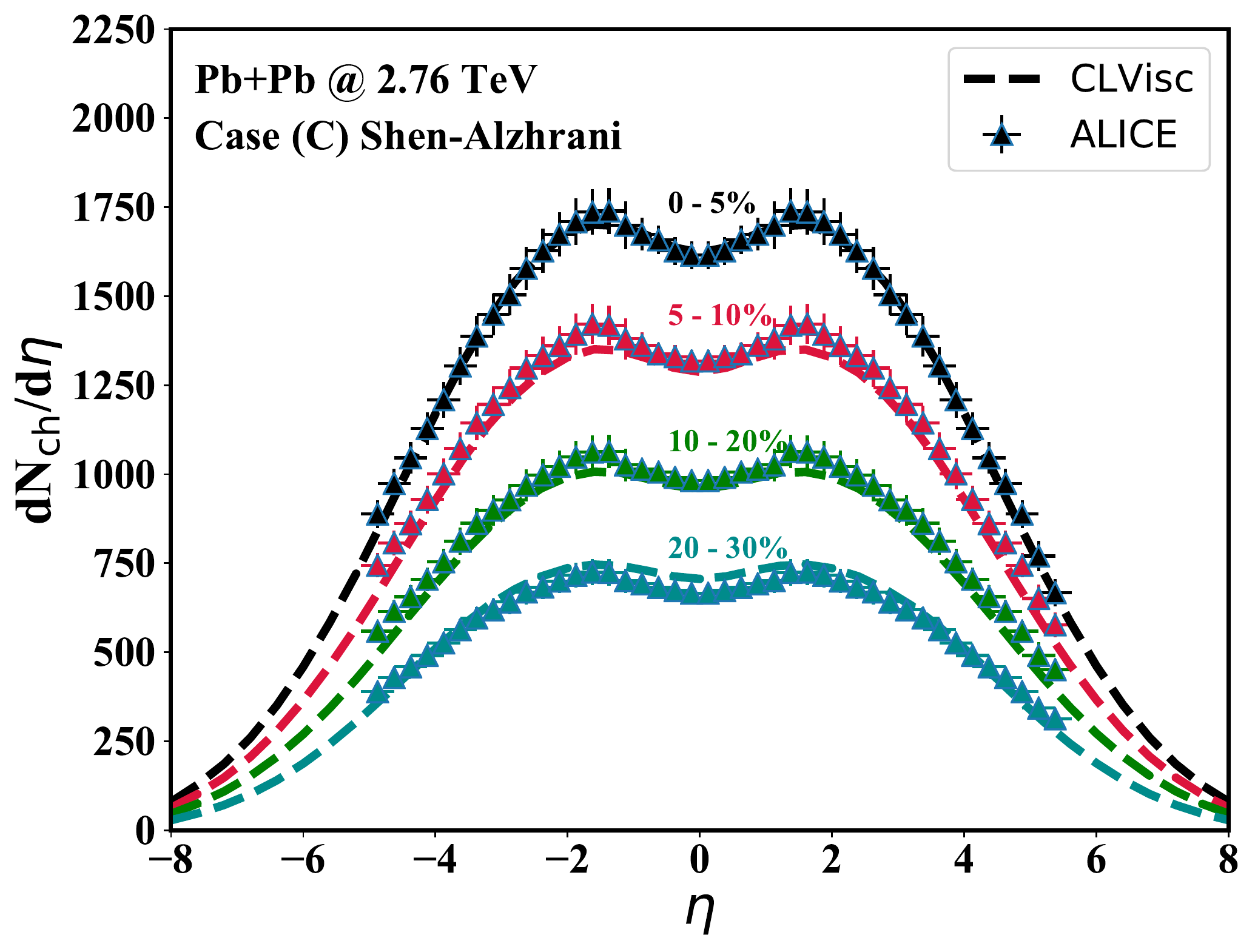}~
\end{center}
\caption{(Color online) Pseudorapidity distribution of charged particles in Pb+Pb collisions at $\snn = 2.76$~TeV in 0-5\%, 5-10\%, 10-20\% and 20-30\% centrality classes, compared between the CLVisc hydrodynamic calculation with three initial condition setups and the ALICE data~\cite{Adam:2015kda}.}
\label{f:pbpb2760dedeta}
\end{figure}

To start with, we validate our model setup by comparing to the pseudorapidity distribution of the final-state charged particles in Fig.~\ref{f:auau200dndeta} (for RHIC) and Fig.~\ref{f:pbpb2760dedeta} (for LHC). As discussed earlier, the model parameters summarized in Tab.~\ref{t:modelparameters} are adjusted to describe these charged particle distributions in the most central collisions at RHIC and LHC. As shown in the figures, after the hydrodynamic parameters are fixed for 0-6\% Au+Au collisions at $\sqrt{s_\text{NN}}=200$~GeV and 0-5\% Pb+Pb collisions at $\sqrt{s_\text{NN}}=2.76$~TeV, our calculation provides reasonable descriptions of the PHOBOS data~\cite{Alver:2010ck} on the $dN_\text{ch}/d\eta$ distributions in other centralities at RHIC and the ALICE data~\cite{Adam:2015kda} at LHC. The same set of parameters in Tab.~\ref{t:modelparameters} can be applied to the three initial condition models under investigation. This provides a reliable baseline for our further study of the directed flow coefficient.

\begin{figure}[!tbp]
\begin{center}
\includegraphics[width=0.75\linewidth]{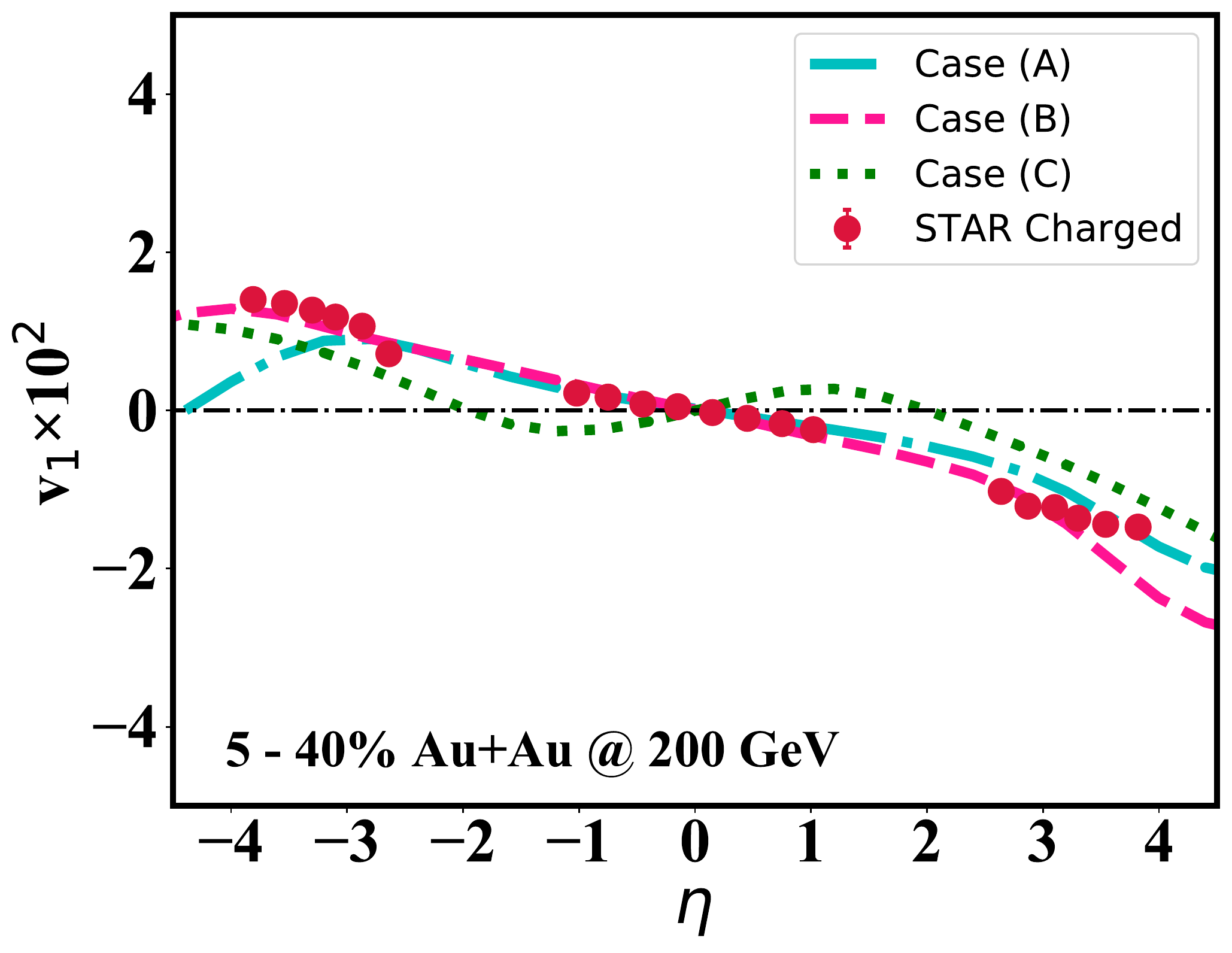}~\\
\includegraphics[width=0.75\linewidth]{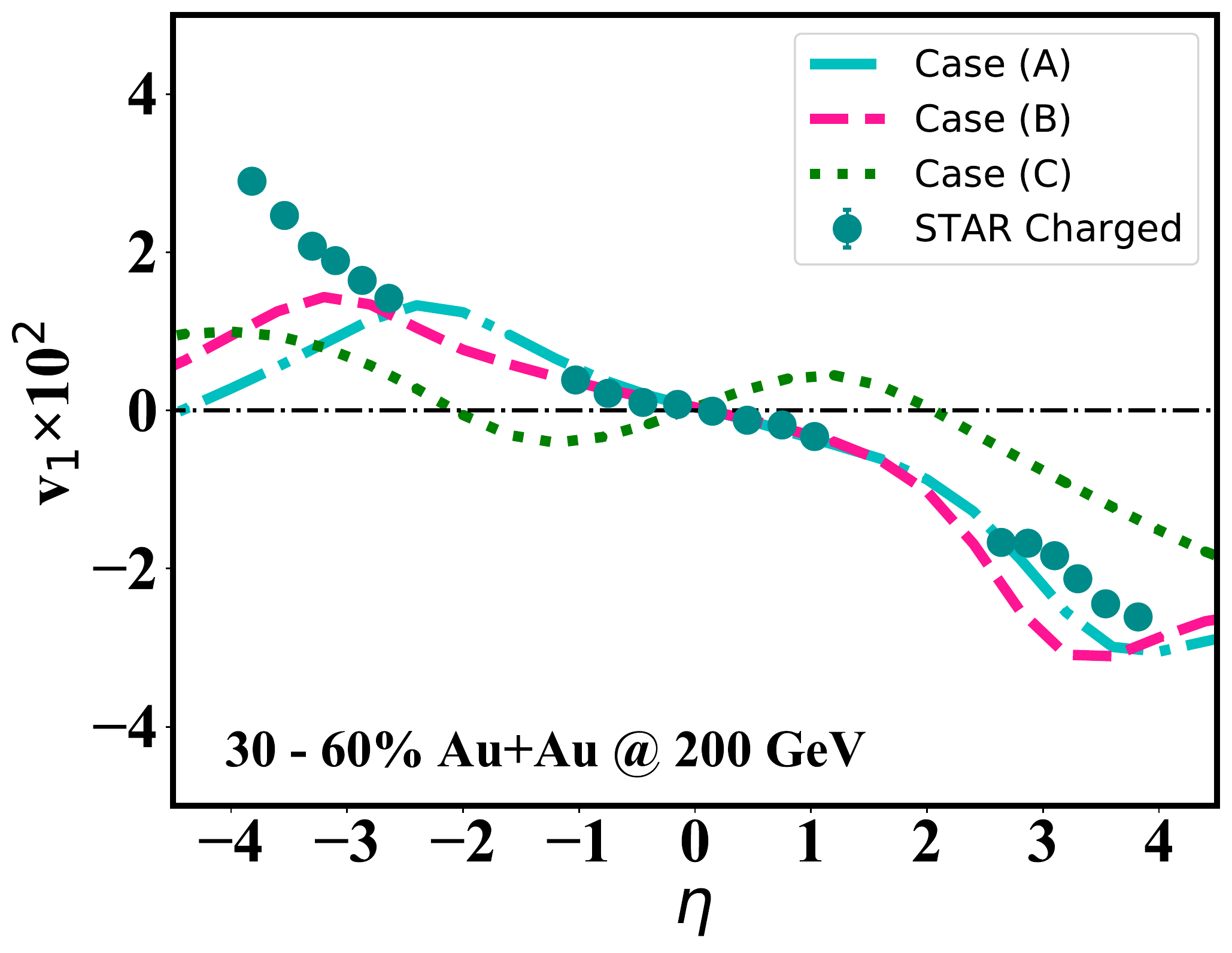}~
\end{center}
\caption{(Color online) Pseudorapidity dependence of the directed flow coefficient in 5-40\% (upper panel) and 30-60\% (lower panel) Au+Au collisions at $\snn = 200$~GeV, compared between the CLVisc hydrodynamic calculation with three initial condition setups and the STAR data~\cite{STAR:2004jwm,Abelev:2008jga}.}
\label{f:v1-200}
\end{figure}

\begin{figure}[!tbp]
\begin{center}
\includegraphics[width=0.75\linewidth]{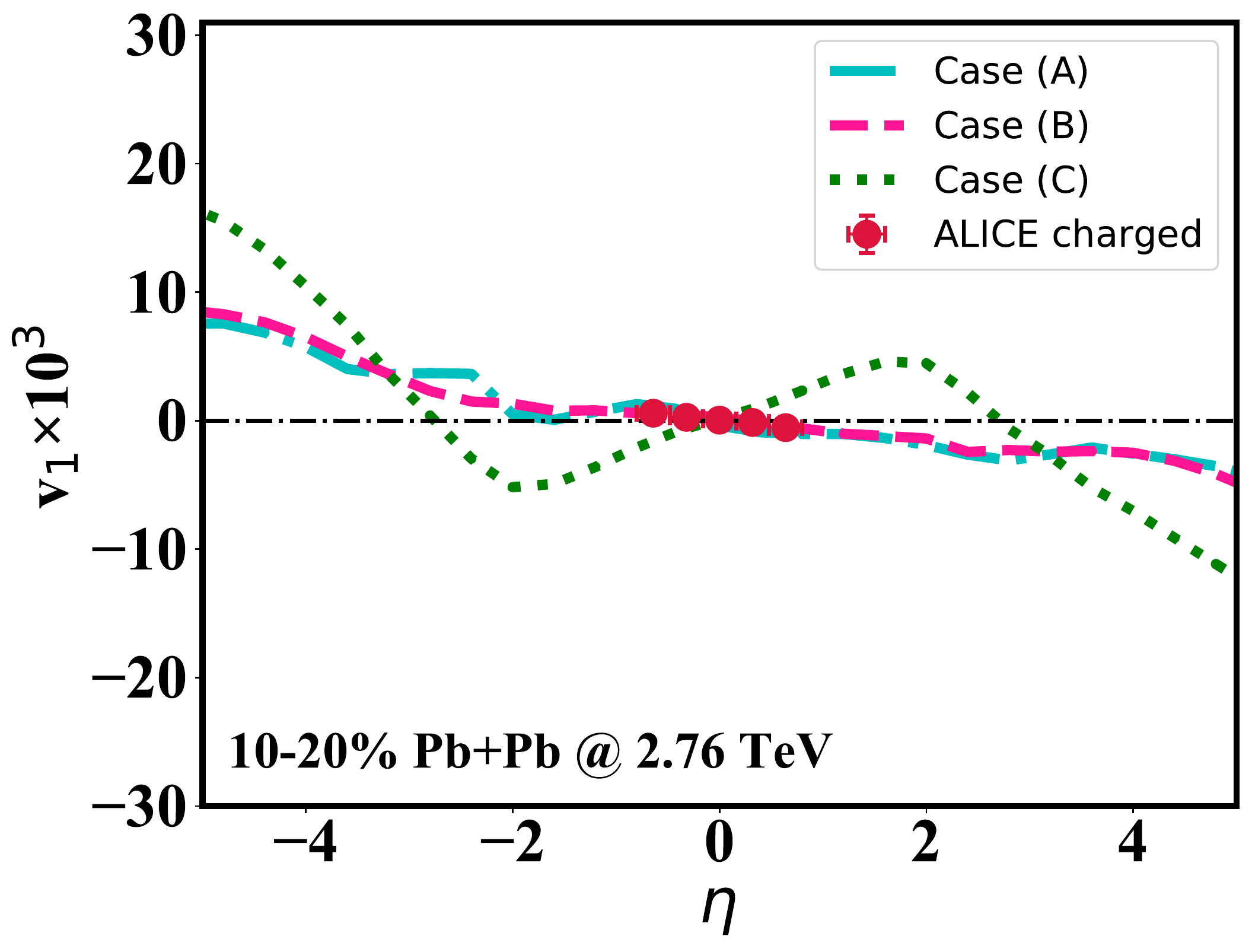}~\\
\includegraphics[width=0.75\linewidth]{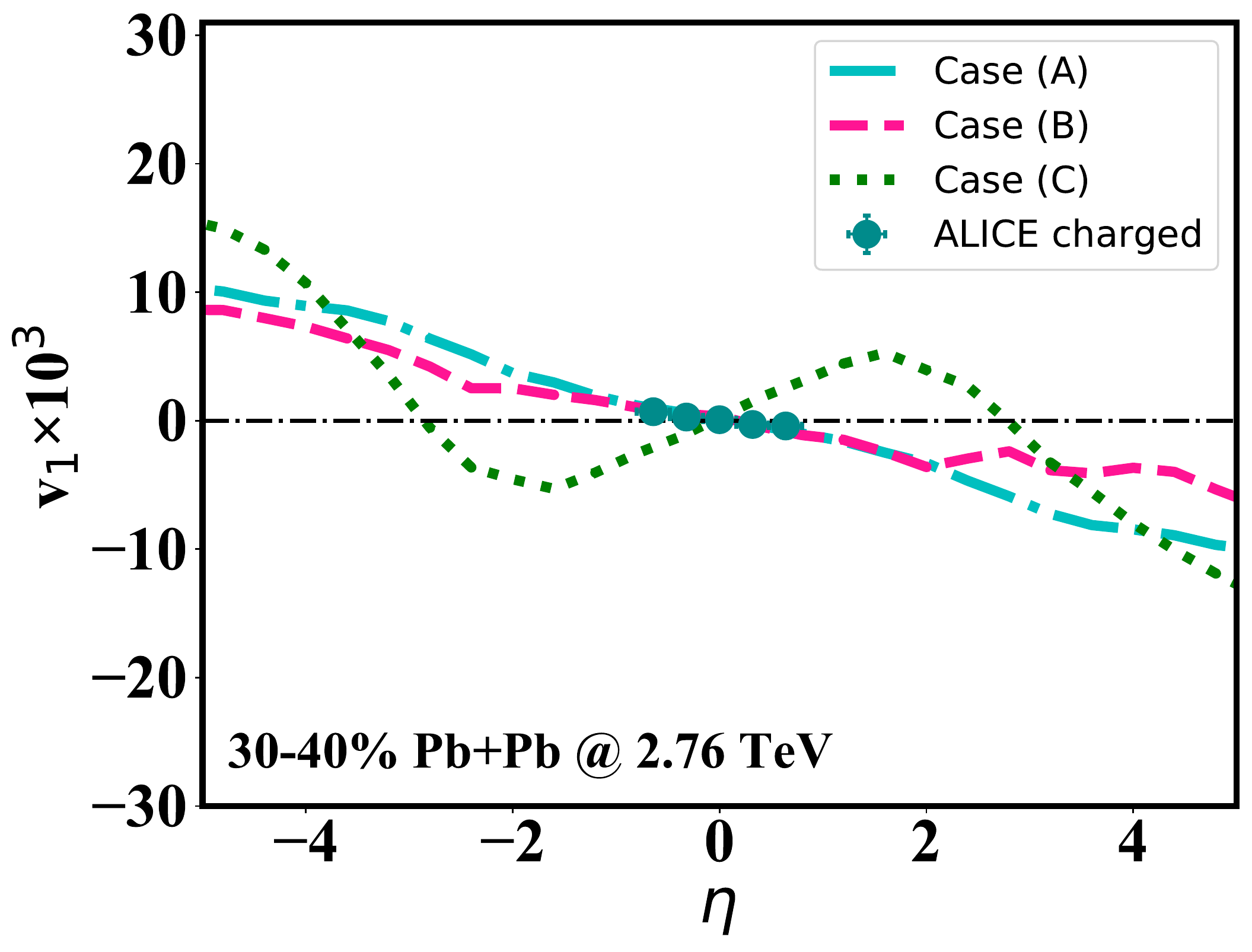}~
\end{center}
\caption{(Color online) Pseudorapidity dependence of the directed flow coefficient in 10-20\% (upper panel) and 30-40\% (lower panel) Pb+Pb collisions at $\snn = 2.76$~TeV, compared between the CLVisc hydrodynamic calculation with three initial condition setups and the ALICE data~\cite{Abelev:2013cva}.}
\label{f:v1-2760}
\end{figure}

In the end, we present results on the charged particle directed flow $v_1$ as functions of pseudorapidity.
Here $v_{1}(\eta)$ is calculated via,
\begin{equation}
\begin{aligned}
v_{1}(\eta)=\langle\cos(\phi-\Psi_{1})\rangle=\frac{\int\cos(\phi-\Psi_{1})\frac{dN}{d\eta d\phi}d\phi}{\int\frac{dN}{d\eta d\phi}d\phi},
\label{eq:v1}
\end{aligned}
\end{equation}
where $\Psi_{1}$ is the first-order event plane of the collision~\cite{Bozek:2010bi}. Since we use the optical Glauber model (described in Sec.~\ref{v1subsect2}) to initialize the energy density distribution of the QGP, event-by-event fluctuations are ignored in this work. As a result, the event plane here should be the same as the spectator plane determined using the deflected neutrons in realistic experimental measurements. A more consistent analysis should be conducted in our future work after event-by-event fluctuations are introduced.
The charged particle $v_1$ is sensitive to the  deformation of the initial energy density, which is governed by $\eta_m$ in the Boz$\dot{\textrm{e}}$k-Wyskiel parametrization, $H_t$ in the CCNU parametrization and $f$ in the Shen-Alzhrani parametrization, as described in Sec.~\ref{v1subsect2}.

In Case (A), $\eta_{m}$ determines the forward-backward correlation length~\cite{Bozek:2010bi,Inghirami:2019mkc}. A larger value of $\eta_{m}$ in Eq.~(\ref{eq:wnpl}) yields a weaker tilt of the initial energy density distribution thus a smaller value of charged particle $v_{1}$. In Case (B), an alternative way of constructing the tilted initial condition is proposed. A larger value of $H_{t}$ in Eq.~(\ref{eq:mnccnu}) results in a more tilted initial energy distribution and therefore a larger magnitude of $v_{1}$ in the end. In Case (C), the parameter $f$ varies the longitudinal extent of the fireball and affects the net longitudinal momentum of the hydrodynamic fields~\cite{Ryu:2021lnx}. For a given collision system in a given centrality, a larger $f$ leads to a smaller magnitude of $v_{1}$, which is constrained by experimental data. Values of these three parameters used in our calculation are summarized in Tab.~\ref{t:para2}.
These parameters only affect the deformation of the medium geometry, but have very weak impact on the $dN_\text{ch}/d\eta$ distributions that have already been fixed by parameters in Tab.~\ref{t:modelparameters}.

\begin{table}[!h]
\begin{center}
\begin{tabular}{ |c| c |c| c| c | }
\hline
           & 200 GeV  & 200 GeV     & 2.76 TeV      &  2.76 TeV        \\
           & Au+Au   & Au+Au     & Pb+Pb       & Pb+Pb         \\
           &  5-40\%        &  30-60\%    &  10-20\%         &  30-40\%        \\
\hline
$b$~(fm)~\cite{Loizides:2017ack,Pang:2018zzo}             & 6.69        & 9.70   & 6.50 & 9.40            \\
\hline
$\eta_{m}$      & 2.80       & 2.10    & 12.10 & 12.00              \\
\hline
$H_{t}$         & 2.90       & 4.30    & 0.70  & 0.70        \\
\hline
$f$~\cite{Ryu:2021lnx}             & 0.15  & 0.15 & 0.05
  & 0.05      \\
\hline
\end{tabular}
\caption{\label{t:para2} Parameters used in different initial condition models for describing the directed flow coefficient measured at RHIC and LHC. The average impact parameters ($b$) are used for different centrality bins.}
\end{center}
\end{table}

Using these parameters, we show the charged particle $v_1$ in Au+Au collisions at $\snn = 200$~GeV in Fig.~\ref{f:v1-200}, upper panel for 5-40\% centrality and lower for 30-60\%. We observe parametrizations in our Case (A) and (B) result in comparable charged particle $v_1$ within $-2 < \eta < 2$, which is also consistent with the STAR data. As expected, the sign of the charged particle $v_1$ is consistent with that of the average flow velocity $v_x$ of the QGP fluid -- positive/negative at backward/forward rapidity. Similar findings for Case (A) and (B) can be observed in Fig.~\ref{f:v1-2760} for the charged particle $v_1$ in Pb+Pb collisions at $\snn = 2.76$~TeV, upper panel for 10-20\% and lower for 30-40\%.

For Case (C), within our current hydrodynamic framework, we are unable to describe the charged particle $v_1$ at either RHIC or LHC by adjusting its $f$ parameter. In the present study, we use $f=0.15$ at RHIC and $f=0.05$ at LHC as suggested by the $\Lambda$ polarization calculation~\cite{Ryu:2021lnx,Shen:private}. If one increases the value of $f$, the slope of $v_1$ {\it vs.} $\eta$ will increase near the mid-pseudorapidity region and further deviate from the experimental data. This implies the importance of the tilted deformation of the initial energy density distribution in understanding the $\eta$-dependence of $v_1$ observed in experiments. Our results for Case (C) here are qualitatively consistent with previous studies~\cite{Shen:2020jwv,Bozek:2010bi}. However, as suggested by Refs.~\cite{Shen:2020jwv,Ryu:2021lnx}, introducing net baryon current into hydrodynamic simulation, varying the $\eta_{w}$ and $\sigma_\eta$ parameters in Eq.~(\ref{eq:Nxy}), or taking into account hadronic rescatterings after the QGP phase may further affect the charged particle $v_1$, or even flip its sign in the mid-$\eta$ region. These effects are not the focus of our present study on the medium geometry effects and will be left for a future investigation.

In the present study, the directed flow coefficients above are analyzed with soft hadrons within $0<p_\mathrm{T}<3.5$~GeV. Since we use the smooth initial condition from the optical Glauber model, our discussion is restricted to the rapidity-odd component of $v_1$ here. The rapidity-even component, including its non-trivial $p_\mathrm{T}$ dependence even at mid-rapidity~\cite{Teaney:2010vd,Luzum:2010fb,Gale:2012rq} is beyond the scope of this work.




\section{Summary and outlook}
\label{v1section4}

In this work, we have performed a systematic study on how the geometry of the initial energy density profile affects the charged particle directed flow in high-energy nuclear collisions. Three different parametrization setups -- Case (A) Boz$\dot{\textrm{e}}$k-Wyskiel, Case (B) CCNU and Case (C) Shen-Alzhrani -- are compared for the initial energy density distribution, and their subsequent time evolutions are simulated using the (3+1)-D viscous hydrodynamic model CLVisc.


Within this framework, we have found the counter-clockwise tilt of the initial energy density profile in the $x$-$\eta_s$ plane, as generated by our Case (A) and (B), yields an increasing/decreasing average pressure gradient $-\langle \partial_x P\rangle$ from zero with respect to time at backward/forward rapidity. The magnitude of this pressure gradient appears larger at larger $|\eta_s|$. This further leads to a negative slope of the average QGP flow velocity $\langle v_x \rangle$ with respect to the space-time rapidity within $|\eta_s|<2$, and in the end the same behavior of $v_1$ {\it vs.} $\eta$ of the final-state charged particles emitted from the QGP. In contrast, without the tilted deformation in the central space-time rapidity region, the shifted initial geometry along the longitudinal direction alone in Case (C) results in the opposite sign of the $x$-component of the average pressure gradient, therefore the average QGP flow velocity and in the end the charged particle $v_1$ at mid-$\eta_s$ (or $\eta$). At large space-time rapidity, the stronger deformation from Case (C) can also generate similar time and $\eta_s$ ($\eta$) dependence of the above quantities as Case (A) and (B). A comparison to the RHIC and LHC data indicates the essential role of the tilted initial energy density profile in helping understand the observed charged particle $v_1$. We note that the correct sign of $-\langle \partial_x P\rangle$ at backward/forward rapidity is the key that drives the final directed flow of soft hadrons. This is more important than the detailed form of how to parametrize the initial condition. Alternative parametrization methods other than the tilted profile may exist, which lead to the similar $\eta_s$ dependence of the pressure gradient.

Our study constitutes a step forward in revealing the source of the directed flow generated in relativistic heavy-ion collisions. Nevertheless, apart from the titled deformation of the initial geometry, other sources exist for the development of directed flow. For instance, (1) deviations from the Bjorken flow in the initial condition could provide additional contribution to the directed flow~\cite{Ryu:2021lnx}. In particular, they could affect the flow pattern of baryons and thus the bulk medium profile~\cite{Bozek:2010bi}. (2) The strong electromagnetic field produced in the early stage of non-central heavy-ion collisions results in deflection of charged particles and influence the charged particle $v_1$~\cite{Inghirami:2019mkc,Gursoy:2014aka,Gursoy:2018yai}, although this effect is suggested small compared to the titled initial geometry~\cite{Inghirami:2019mkc}. (3) The charged particle $v_1$ can also be contributed by the initial decelerated baryons and hadronic cascade after the QGP phase, especially at lower collision energy~\cite{Shen:2020jwv,Ryu:2021lnx}. Moreover, a combination of hydrodynamic model and afterburner hadronic transport, such as UrQMD~\cite{Bass:1998ca,Zhao:2021vmu} and SMASH~\cite{Petersen:2018jag,Wu:2021fjf}, are found necessary for more stringent constraints on the initial stage of nuclear collisions. These should be incorporated in our future calculation for a more precise understanding of the directed flow.

\begin{acknowledgements}
We are grateful for helpful discussions with Chun Shen, Jiaxing Zhao, Guang-You Qin and Long-Gang Pang. This work was supported by the National Natural Science Foundation of China (NSFC) under Grant Nos.~11935007, 12175122 and 2021-867, Guangdong Major Project of Basic and Applied Basic Research No.~2020B0301030008, the Natural Science Foundation of Hubei Province No.~2021CFB272, the Education Department of Hubei Province of China with Young Talents Project No.~Q20212703 and
the Xiaogan Natural Science Foundation under Grant No.~XGKJ2021010016. Computational resources were provided by the Center of Scientific Computing
at the Department of Physics and Electronic-Information Engineering, Hubei Engineering University.
\end{acknowledgements}

\bibliographystyle{unsrt}
\bibliography{v1ref}

\end{document}